\documentclass[journal]{vgtc}                     

\onlineid{1391}

\vgtccategory{IEEE VIS Paper}

\vgtcpapertype{Representations \& Interaction}

\title{StyleRF-VolVis: Style Transfer of Neural Radiance Fields\\for Expressive Volume Visualization}

\author{%
  \authororcid{Kaiyuan Tang}{0009-0001-3512-0112} and
  \authororcid{Chaoli Wang}{0000-0002-0859-3619}
}

\authorfooter{
  \item K.\ Tang and C.\ Wang are with 
the Department of Computer Science and Engineering, University of Notre Dame, Notre Dame, IN 46556, USA.
E-mail: \{ktang2, chaoli.wang\}@nd.edu.
}

\abstract{In volume visualization, visualization synthesis has attracted much attention due to its ability to generate novel visualizations without following the conventional rendering pipeline. However, existing solutions based on generative adversarial networks often require many training images and take significant training time. Still, issues such as low quality, consistency, and flexibility persist. This paper introduces StyleRF-VolVis, an innovative style transfer framework for expressive volume visualization (VolVis) via neural radiance field (NeRF). The expressiveness of StyleRF-VolVis is upheld by its ability to accurately separate the underlying scene geometry (i.e., content) and color appearance (i.e., style), conveniently modify color, opacity, and lighting of the original rendering while maintaining visual content consistency across the views, and effectively transfer arbitrary styles from reference images to the reconstructed 3D scene. To achieve these, we design a base NeRF model for scene geometry extraction, a palette color network to classify regions of the radiance field for photorealistic editing, and an unrestricted color network to lift the color palette constraint via knowledge distillation for non-photorealistic editing. We demonstrate the superior quality, consistency, and flexibility of StyleRF-VolVis by experimenting with various volume rendering scenes and reference images and comparing StyleRF-VolVis against other image-based (AdaIN), video-based (ReReVST), and NeRF-based (ARF and SNeRF) style rendering solutions.}


\teaser{
\begin{center}
	$\begin{array}{c@{\hspace{0.0in}}c}
	\includegraphics[height=1.125in]{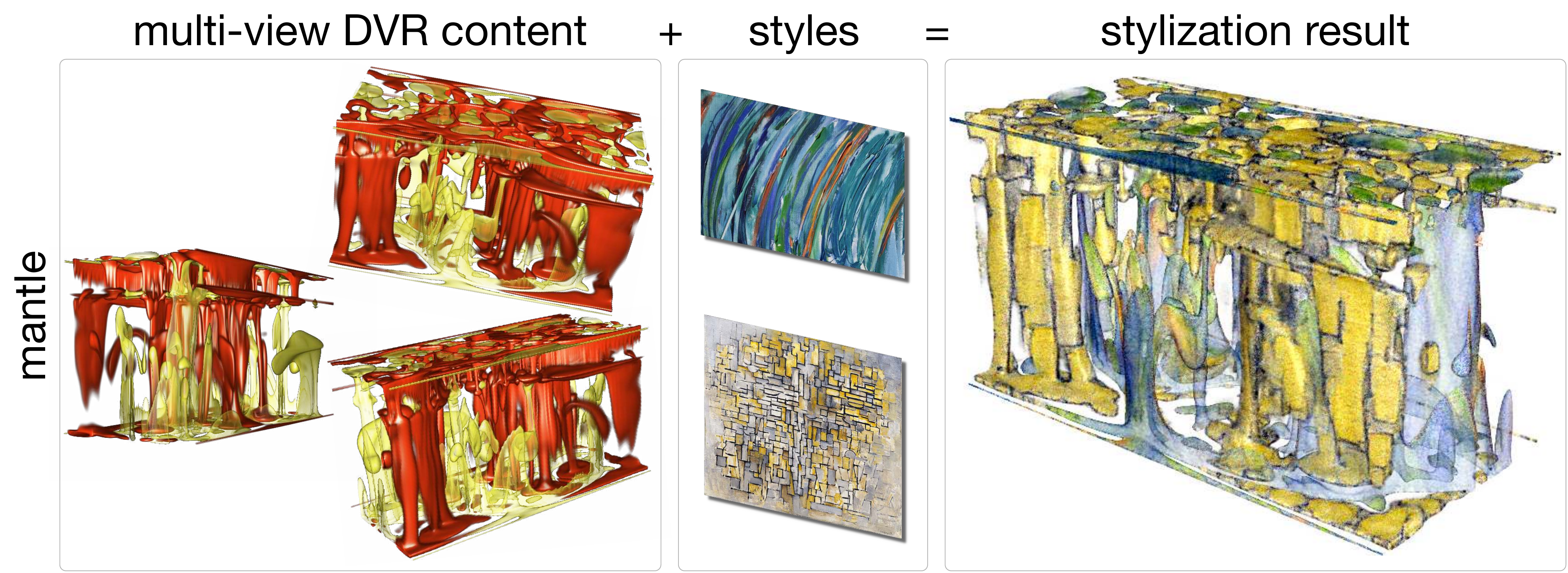} &
	\includegraphics[height=1.125in]{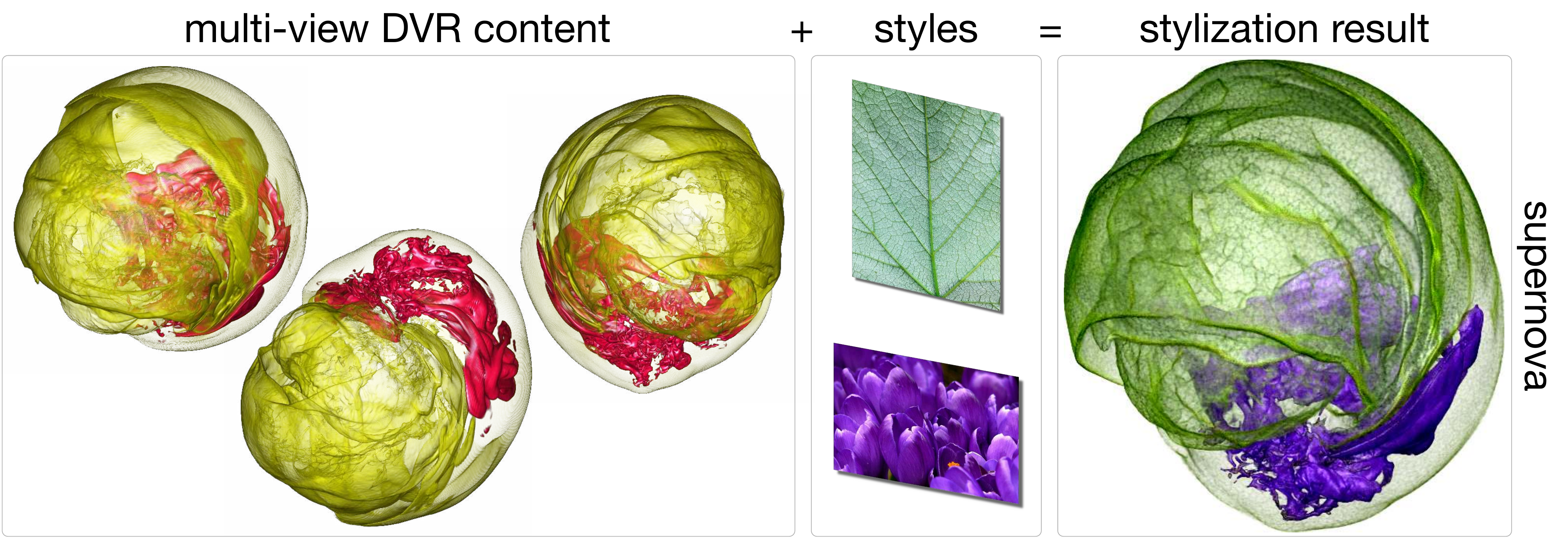} \\
	\end{array}$
\end{center}
\vspace{-0.275in}
\caption{Given a set of multi-view DVR images produced from a volumetric dataset, StyleRF-VolVis transfers arbitrary styles from reference images to the reconstructed 3D scene, synthesizing high-quality novel views with visual content consistency.}
  \label{fig:teaser}
}

\graphicspath{{figs/}{figures/}{pictures/}{images/}{./}} 

\usepackage{tabu}                      
\usepackage{booktabs}                  
\usepackage{lipsum}                    
\usepackage{mwe}                       

\usepackage{mathptmx}                  
\usepackage{multirow}
\usepackage{graphicx}
\usepackage{times}
\usepackage[linesnumbered,ruled]{algorithm2e}
\usepackage{amsfonts}
\usepackage{booktabs}         
\usepackage{gensymb}
\usepackage{amsmath}

\usepackage{comment}
\usepackage{times} 
\usepackage{caption}
\usepackage{bm}
\usepackage{multirow}
\usepackage{cclicenses}
\usepackage{makecell}
\usepackage{lscape,array}
\usepackage{color} 
\usepackage{anyfontsize}
\usepackage{soul}
\usepackage{verbatim}
\usepackage{url}
\usepackage{float}

\DeclareMathOperator*{\base}{BASE}
\DeclareMathOperator*{\convex}{CONVEX}
\DeclareMathOperator*{\palette}{PALETTE}
\DeclareMathOperator*{\volume}{VOLUME}
\DeclareMathOperator*{\colors}{COLOR}
\DeclareMathOperator*{\nnfm}{NNFM}
\DeclareMathOperator*{\lin}{lin}
\SetKwInput{KwInput}{Input}                
\SetKwInput{KwOutput}{Output}              
\definecolor{commentcolor}{RGB}{110,154,155}   

\newcommand{\pin}[1]{{\color{black} #1}}

\newenvironment{myitemize}{
\begin{itemize}
 \setlength{\itemsep}{1pt}
 \setlength{\parskip}{0pt}
 \setlength{\parsep}{0pt}}{\end{itemize}
 
}

\usepackage[switch,pagewise]{lineno} 

\begin{document}


\firstsection{Introduction}

\maketitle

Since the early 1990s, volume visualization (VolVis) has been a central topic in scientific visualization research. One of the most popular techniques for VolVis is direct volume rendering (DVR). It works by casting rays from the image plane to the 3D volume, gathering samples along each ray, mapping them to visual quantities (i.e., colors and opacities) via transfer function (TF) lookups, and compositing the final color for each pixel in the rendering image. Thanks to the astonishing advances in graphics hardware, DVR can be efficiently implemented to achieve interactive framerates and high-quality visualizations, providing superior capability for users to explore volumetric datasets interactively. This practice has been the norm for the past two decades. 

The recent surge of deep learning for scientific visualization research~\cite{Wang-DL4SciVis} has sprouted new opportunities for VolVis. Leveraging the capabilities of generative adversarial networks (GANs), one can train a network to synthesize rendering results under novel viewpoints, TFs, or other parameters, eliminating the need to access the original volumetric data and bypassing the conventional rendering pipeline~\cite{Berger-TVCG19, Hong-DNN-VolVis}. These seemingly impossible advances could have shocked many researchers just several years ago but are widely understood by the research community nowadays. After all, generative AI has swept across many fields, culminating in its extraordinary power to synthesize novel images and videos from text prompts and level the playing field for the masses.  

Inspired by this fantastic trend, in this paper, we aim to design a deep learning framework to accomplish {\em expressive visualization} via {\em style transfer} for VolVis. 
Similar to prior work~\cite{Berger-TVCG19, Hong-DNN-VolVis, He-InsituNet, Han-TVCG23}, we are supposed to be given a set of 2D images captured from the sample viewpoints and rendered with \pin{a fixed, reasonably good, yet unknown} TF. Also, we assume no access to the 3D volumetric dataset during training or inference. 
We aspire to develop a new model that achieves expressive visualization by meeting the following goals ({\bf G1} to {\bf G3}) for end users. 
First, they can freely explore the DVR scene in excellent visual quality from previously unseen viewpoints ({\bf G1}).
Second, they can further flexibly recolor the rendering results {\em photo-realistically} by editing the underlying colors, opacities, and lighting effects ({\bf G2}). 
Third, and most importantly, they can even intuitively modify the rendering results {\em non-photorealistically} by transferring styles (e.g., salient brushstrokes) from any image or painting of interest to the DVR scene ({\bf G3}). 

Realizing {\bf G1} is relatively straightforward. 
Unlike GAN-based techniques, state-of-the-art solutions based on {\em neural radiance field} (NeRF) can deliver high-resolution, high-quality novel view synthesis results using a set of sample images. 
Nevertheless, significant challenges exist for achieving {\bf G2} and {\bf G3} due to the considerable gaps between the given 2D images and the 3D scene we aim to reconstruct and edit. 
For {\bf G2}, the main challenge lies in accurately extracting the contributing color, opacity, and lighting information for faithful downstream edits to ensure visual content consistency. 
For {\bf G3}, the grand challenge is to move beyond a fixed number of colors decoded from the original rendering images and adapt to abundantly rich color patterns or textures in the reference images. 

We introduce StyleRF-VolVis, achieving style transfer of NeRFs for expressive VolVis. 
\pin{For large volumetric data, NeRF-based scene representations are space efficient and thus have implications for altering renderings in a compressive manner.} 
\pin{At its core, we highlight three key components of StyleRF-VolVis to realize the stated goals ({\bf G1} to {\bf G3}).} 
First, we employ a {\em base NeRF model} to learn the density representation (i.e., content) of the 3D volumetric data from the collection of 2D training images ({\bf G1}). 
\pin{This component accurately extracts the underlying scene geometry, paving the way for successful subsequent style editing.}
Second, we design a new {\em palette color network} that classifies regions of the radiance field (RF) and extracts a color palette (i.e., appearance) from multi-view images ({\bf G2}). 
\pin{This component enables {\em photorealistic style editing} by modifying the original rendering's color, opacity, and lighting while maintaining visual content consistency across the views.}
Third, to remove the color palette constraint, we propose a novel {\em knowledge distillation} solution that transfers the color information from the palette color network (i.e., {\em teacher model}) to an {\em unrestricted color network} (i.e., {\em student model}) ({\bf G3}). 
\pin{This component allows users to assimilate a wide spectrum of color patterns and textures from any reference image, supporting expressive visualization of the DVR scene via {\em non-photorealistic style editing}.} 
Figure~\ref{fig:teaser} highlights the capability of StyleRF-VolVis on two datasets, each using two styles extracted from different reference images. 
The contributions of our work are:
\begin{myitemize}
\item \pin{We revisit style transfer for VolVis by presenting an innovative NeRF-based solution that supports expressive photorealistic and non-photorealistic style editing.}
\item StyleRF-VolVis represents a significant leap forward for visualization synthesis of volumetric data, advancing the state-of-the-art solutions in quality, consistency, and flexibility. 
\item \pin{We show the consistency and flexibility of StyleRF-VolVis on various combinations of DVR scenes and reference images.}
\item We compare StyleRF-VolVis with other image-based (AdaIN), video-based (ReReVST), and NeRF-based (ARF and SNeRF) style rendering solutions to demonstrate its superior quality. 
\end{myitemize}

\vspace{-0.1in}
\section{Related work}

This section discusses related work on deep learning for VolVis, style transfer, NeRF for stylization, and knowledge distillation.

{\bf Deep learning for VolVis.}
Over the past few years, deep learning has emerged as a promising solution for improving the DVR process. 
One application is to use deep learning models to replace the traditional DVR pipeline. 
For instance, Berger et al.\ \cite{Berger-TVCG19} constructed a GAN to synthesize volume-rendered images by investigating the image space of volume rendering under various TFs and viewing parameters. 
Hong et al.\ \cite{Hong-DNN-VolVis} proposed a GAN to synthesize high-resolution images from volume data under the desired rendering effect without knowing the TF. 
He et al.\ \cite{He-InsituNet} developed InsituNet, a surrogate model that correlates simulation and visualization parameters with visualization outcomes, allowing users to preview visualization results under different simulation settings with a trained model. 
Shi et al.\ \cite{Shi-VDLSurrogate} built a view-dependent neural-network-latent-based surrogate model that supports producing high-resolution visualization results. 
Han and Wang~\cite{Han-TVCG23} presented CoordNet, a coordinate-based neural network to synthesize rendering images under novel viewpoints given a set of DVR images for training. 
Another application employs a scene representation network (SRN) to represent volumetric data, minimizing disk storage requirements and enabling interactive neural rendering without direct access to volume data. 
For example, Weiss et al.\ \cite{Weiss-fVSRN} designed a dense-grid encoding method fV-SRN that directly renders from compressed representation with no additional memory for storing volume data during rendering. 
Wu et al.\ \cite{Wu-MHT} leveraged multiresolution hash grid encoding and achieved fast volume encoding and real-time rendering. 
Wurster et al.\ \cite{Wurster-APMGSRN} developed APMGSRN, an adaptively placed multi-grid SRN with a domain decomposition training approach for efficient VolVis. 
Other deep learning works in the context of VolVis focus on data or visualization generation~\cite{Han-VIS19,Han-TVCG22,Han-VIS21,Tang-CG24,Weiss-TVCG21}, compression~\cite{Lu-CGF21,Tang-PVIS24,Gu-CG23,YF-Lu-VISSP24}, translation~\cite{Han-VIS20,Yao-CG23}, and completion~\cite{Han-VI22}. 
Our work represents the first step in applying style transfer techniques to volume rendering results using a NeRF model. Compared with existing works~\cite{Berger-TVCG19, Hong-DNN-VolVis, He-InsituNet, Han-TVCG23}, StyleRF-VolVis takes a smaller set of DVR images for training and infers higher-quality results, maintaining content consistency across different viewpoints and supporting flexible style transfer. 

{\bf Style transfer.}
Ever since Gatys et al.\ \cite{Gatys-CVPR16} demonstrated the capability of convolutional neural networks (CNN) in applying diverse artistic styles to natural images, neural style transfer~\cite{Jing-TVCG20} has emerged as a trending topic in both academia and industry. Numerous techniques have been developed to improve or extend the original algorithm.
Johnson et al.\ \cite{Johnson-ECCV16} introduced a feed-forward network to solve the slow optimization problem and achieved real-time style transfer. 
Huang et al.\ \cite{Huang-ICCV17} further maintained speed comparable to~\cite{Johnson-ECCV16} without sacrificing the flexibility of transferring inputs to arbitrary new styles. This was realized using an adaptive instance normalization layer that aligns the statistics information of the content and style features. 
Ruder et al.\ \cite{Ruder-GCPR16} extended style transfer to video sequences by computing optical flow to achieve consistent style transfer across frames. 
Wang et al.\ \cite{Wang-ReReVST} presented ReReVST that relaxes the objective function of style loss to make the transfer more robust to motions in content video. 

In VolVis, Bruckner and Gr{\"o}ller~\cite{Brucker-CGF07} introduced a style TF using the lit sphere, which assigns the optical properties of voxel values with rendering styles instead of simple colors and opacities. 
However, extracting or designing the lit sphere requires artist involvement, and the rendering relies on a traditional pipeline. Our StyleRF-VolVis adopts an advanced NeRF representation, supporting end-to-end style transfer and rendering without accessing the original volume. 

{\bf NeRF for stylization.}
Since the groundbreaking work of Mildenhall et al.\ \cite{Mildenhall-NeRF} in 2020, NeRF has been widely used for novel view synthesis. 
\pin{We refer readers to recent survey papers~\cite{NeRFSurvey-CGF22, NeRFStyleSurvey} to follow the roadmap of NeRF-based applications.}
While extensive studies~\cite{Barron-ICCV21, Fridovich-Keil-CVPR22, Thomas-InstantNGP, Kerbl-3DGS, Hu-ICCV23} emphasize the quality or speed enhancement of NeRF, there is also a growing body of work focusing on editing a base NeRF to perform 3D style transfer of a scene. 
Generally, editing NeRF for style transfer can be classified into {\em photorealistic} and {\em non-photorealistic}.
One main task for {\em photorealistic editing} is recoloring the scene. 
For example, Kuang et al.\ \cite{Kuang-PaletteNeRF} and Gong et al.\ \cite{Gong-RecolorNeRF} utilized optimizable base colors in a palette to fit the scene and modify the base colors to recolor the scene during inference.
Lee et al.\ \cite{Lee-ICCV23} modified the essential weights in the color multilayer perceptron (MLP) of a trained NeRF to perform local recoloring. 
Unlike photorealistic editing, {\em non-photorealistic editing} transfers more abstract information, such as textures or brushstrokes of an artistic work. 
For instance, Chiang et al.\ \cite{Chiang-WACV22} utilized a hypernetwork to transfer style features into the NeRF representation. 
Huang et al.\ \cite{Huang-CVPR22} proposed a mutual learning strategy to integrate 2D stylization with NeRF geometry consistency. 
Zhang et al.\ \cite{Zhang-ARF} designed a nearest neighbor-based loss to stylize a trained NeRF, which performs better stylization quality compared with standard Gram matrix-based loss. 
Liu et al.\ \cite{Liu-CVPR23} transformed the grid features within RFs to match a reference style. Even though the quality is less impressive than~\cite{Zhang-ARF}, it supports zero-shot transfer to an unseen reference image.
StyleRF-VolVis differs from all these works in that it supports photorealistic and non-photorealistic editing in a single framework. We will show that conducting photorealistic editing before non-photorealistic editing yields better quality and flexibility for stylization results.

\begin{figure*}[!htb]
\centering
\includegraphics[width=1\linewidth]{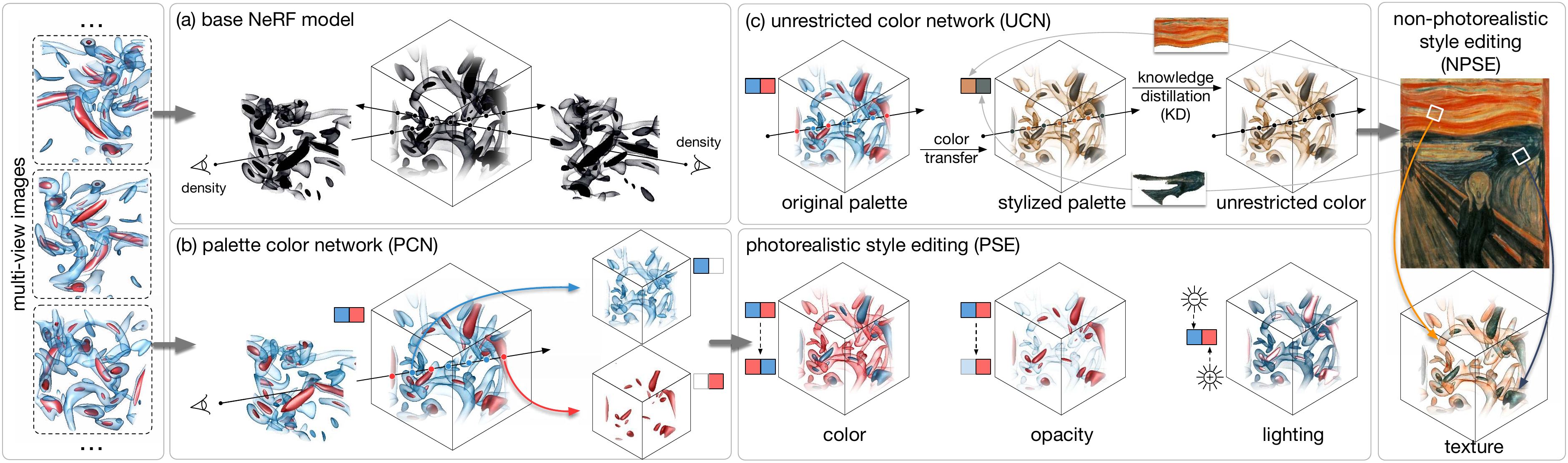}\\
\vspace{-.1in}
\caption{The workflow of StyleRF-VolVis. (a) Given a collection of multi-view images, we first optimize a base NeRF model for accurate density representation. (b) Then, we train a PCN for PSE, allowing color, opacity, and lighting changes. (c) Next, we utilize KD to optimize a UCN with no color palette constraint. Optimized with the stylization loss, the trained UCN can produce NPSE results.}
\label{fig:workflow}
\end{figure*}

{\bf Knowledge distillation.}
Hinton et al.\ \cite{Hinton-arxiv15} introduced {\em knowledge distillation} (KD), which optimizes a small network to match the predictions of a large model. KD has been used in model compression and optimization~\cite{Hinton-arxiv15, Luo-AAAI16, Han-KDINR}. 
In the NeRF space, Reiser et al.\ \cite{Reiser-ICCV21} leveraged a large global pre-trained NeRF to speed up the training of multiple small local NeRFs. 
Barron et al.\ \cite{Barron-CVPR22} improved the rendering quality with an online distillation strategy. 
Wang et al.\ \cite{Wang-ECCV22} designed R2L to learn the light fields of a pre-trained NeRF model effectively. 
Fang et al.\ \cite{Fang-AAAI23} proposed progressive volume distillation, which provides a systematic KD method that allows the distillation between explicit and implicit NeRF architectures. 
Instead of applying KD to optimize a small network for model compression or optimization purposes, we utilize KD to transfer the knowledge of a photo-realistically edited RF to improve the quality of later non-photorealistic editing.


\vspace{-0.05in}
\section{StyleRF-VolVis}
\label{sec:algorithm}

The objective of StyleRF-VolVis is to edit the {\em style} (i.e., appearance) of the DVR scene represented by a NeRF model without losing the original {\em content} (i.e., geometry) information. The style can be {\em photorealistic} (such as adjusting the original rendering's {\em color}, {\em opacity}, and {\em lighting} attributes) or {\em non-photorealistic} (such as altering the {\em texture} to resemble the brushstroke of an artistic image). 
Distinct from conventional 3D neural stylization methods, our StyleRF-VolVis aims to unify two types of style editing into a single framework. 
Additionally, from a visualization perspective, users design TFs to classify the regions of interest in the DVR process. 
These TFs map distinct colors and opacities to each region users specify, highlighting relevant data ranges through visual mapping. 
Such visual content is critical in obtaining helpful visualization. Our method must keep the original visual content consistent across the views to ensure the stylized visualization remains meaningful, facilitating accurate interpretation and analysis.

Figure~\ref{fig:workflow} illustrates the three-stage workflow of StyleRF-VolVis: 
(a) a {\em base NeRF model} for optimizing the density representation of the scene, 
(b) a {\em palette color network} (PCN) for supporting {\em photorealistic style editing} (PSE) and maintaining visual content consistency during style editing, and 
(c) an {\em unrestricted color network} (UCN) for enabling {\em non-photorealistic style editing} (NPSE). 
Accordingly, our method is structured into three optimization stages, each dedicated to training the parameters of one network while others are frozen for stability.

The first stage focuses on training a base NeRF model to reconstruct the scene geometry for subsequent style editing. 
The parameters of the trained base NeRF are then frozen to ensure that the following style editing does not affect the underlying scene geometry.

The second stage optimizes the PCN. 
Similar to TFs that differentiate volumetric regions with user-defined colors and opacities, the goal here is to classify regions of the RF with 
a color palette extracted from multi-view images. 
To this end, the PCN is optimized to fit input images with a weighted linear combination of extracted palette colors, given the density representation obtained from the previous stage.
By learning to represent the colors of the RF, our model achieves PSE, including modifying the original rendering's color, opacity, and lighting.

The number of palette colors inherently restricts the network's ability to utilize more diverse color patterns for NPSE, where each style often consists of multiple colors. 
To address this issue, 
in the third stage, we start with stylizing the palette colors to match the target style colors. 
Leveraging KD, we then transfer the color information from the PCN (i.e., {\em teacher model}) to a UCN (i.e., {\em student model}), representing the colors of the RF without the color palette constraint. 
In addition, we can apply different styles to different visual regions extracted from the PCN.  
As a result, we stylize the visualization scene with NPSE while maintaining the original visual content.

\pin{The rest of Section~\ref{sec:algorithm} is structured as follows. In Section~\ref{subsec:base-nerf}, we provide the preliminary knowledge of NeRF representation~\cite{Mildenhall-NeRF} and how we optimize the base NeRF model. In Sections~\ref{subsec:PCN}~and~\ref{subsec:PSE}, we discuss how we adapt NeRF's palette-based recoloring strategy~\cite{Gong-RecolorNeRF, Kuang-PaletteNeRF, Tojo-CGF22} to fit PCN and achieve PSE. The utilization of KD~\cite{Fang-AAAI23} to extract UCN was described in Section~\ref{subsec:UCN}, followed by the NPSE details that combine stylization loss~\cite{Zhang-ARF, Kolkin-NNST} and unsupervised segmentation techniques~\cite{Kirillov-arxiv23} in Section~\ref{subsec:NPSE}. Finally, we brief our interactive interface design in Section~\ref{subsec:GUI}.}

\vspace{-0.05in}
\subsection{Base NeRF Model} 
\label{subsec:base-nerf}

StyleRF-VolVis is built upon the scene representation from a base NeRF model. 
In general, a NeRF model~\cite{Mildenhall-NeRF} represents the scene with two neural functions: a {\em density function} $\sigma(\textbf{x})$ that maps any 3D position $\textbf{x}$ to a density value $\sigma$ and a {\em color function} $\textbf{c}(\textbf{x}, \textbf{d})$ that outputs RGB color $\textbf{c}$ given an arbitrary 3D position $\textbf{x}$ and viewing direction $\textbf{d}$. 
NeRF samples rays from the camera origin $\textbf{o}$ to the rendering pixels for each training camera view $\textbf{d}$. For one sample ray $\textbf{r}=(\textbf{o},\textbf{d})$, if we sample $M$ points along ray $\textbf{r}$ with 3D positions $\textbf{x}_{1\dots M}$ at depths $\textbf{t}_{1\dots M}$. The predicted pixel color $\hat{\textbf{c}} (\textbf{r})$ of ray $\textbf{r}$ is computed as
\begin{equation}
\label{eqn:baseNeRFColor}
	\hat{\textbf{c}}(\textbf{r})=\sum^{M}_{i=1}\alpha_i(1-\omega_i)\textbf{c}(\textbf{x}_i, \textbf{d}),
\end{equation}
where $\omega_i = \exp(-(\textbf{t}_i-\textbf{t}_{i-1})\sigma(\textbf{x}_i))$ represents the transmittance along the ray between 
$\textbf{x}_i$ and $\textbf{x}_{i-1}$. 
$\alpha_i=\prod^{i-1}_{j=1}\omega_i$ denotes the attenuation of the ray from its origin $\textbf{o}$ to 
$\textbf{x}_i$.
We select the advanced architecture from Instant-NGP~\cite{Thomas-InstantNGP} to optimize the RF, ensuring fast convergence and efficient rendering. Specifically, Instant-NGP comprises a multiresolution hash-grid encoding and a tiny MLP for efficient optimization.
During the optimization of the base NeRF, we optimize with a loss $\mathcal{L}_{\base}$ defined as
\begin{equation}
\label{eqn:baseNeRFLoss}
	\mathcal{L}_{\base} = ||\textbf{c}(\textbf{r}) - \hat{\textbf{c}}(\textbf{r})||^{2}_2,
\end{equation}
where $\textbf{c}(\textbf{r})$ and \pin{$\hat{\textbf{c}}(\textbf{r})$} are the ground-truth (GT) and predicted pixel colors corresponding to ray $\textbf{r}$.
After optimizing the base NeRF, we freeze its model parameters in the subsequent stages to avoid style editing influencing the scene geometry. Note that the colors of the base NeRF will not be used in rendering the stylized RF; we employ the PCN for PSE and the UCN for NPSE. 

\begin{figure}[htb]
\centering
\includegraphics[width=1\linewidth]{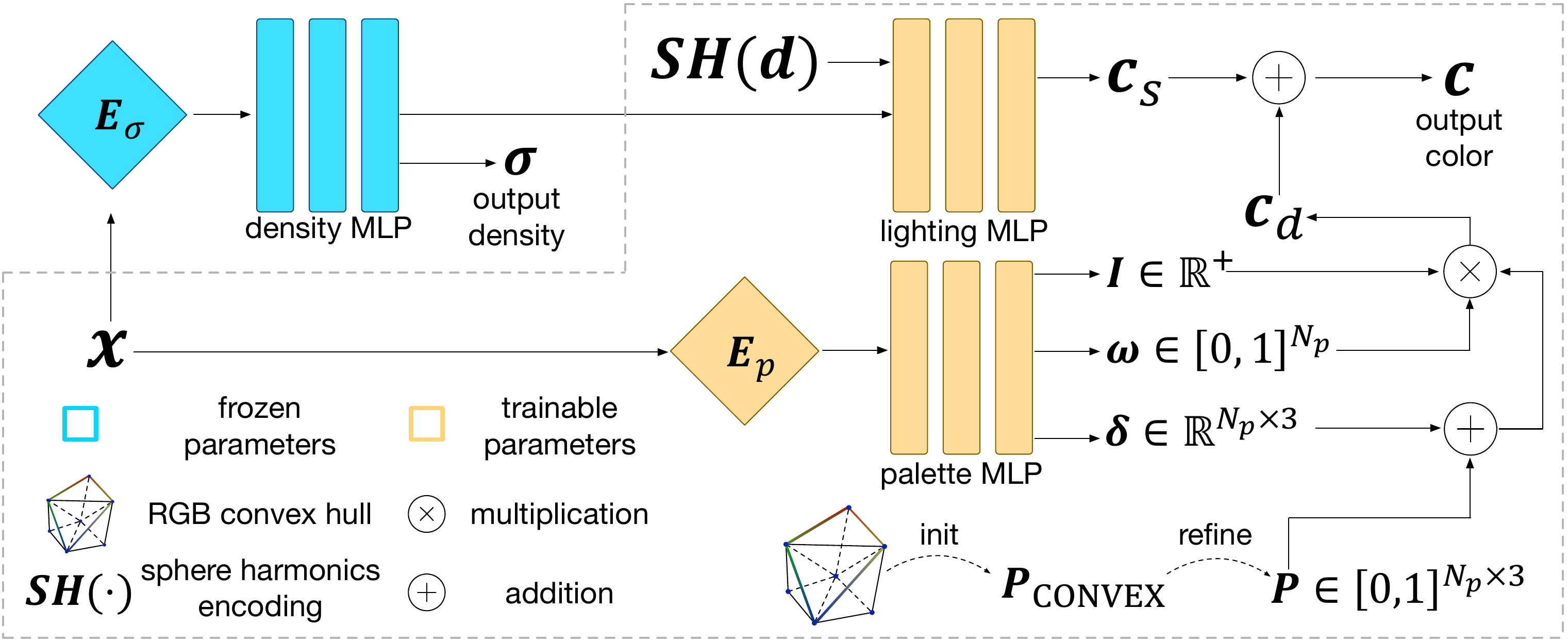}\\
\vspace{-.1in}
\caption{The PCN architecture. 
The final predicted color is obtained by summing 
the view-dependent specular color output from the lighting MLP and
the view-independent diffuse color output from the palette MLP. 
After training, palette weights $\omega$ are used for classifying points within the RF.} 
\label{fig:palette-based}
\end{figure}

\vspace{-0.05in}
\subsection{Palette Color Network (PCN)}
\label{subsec:PCN}

After training the base NeRF model for accurate density representation, we optimize a PCN that represents the scene appearance with 
colors from a palette. 
The trained PCN can classify any 3D position within the RF with its palette color. 
The classification of the PCN preserves region classification from the user-defined TF, ensuring visual content consistency of the downstream style editing.
Figure~\ref{fig:palette-based} shows the PCN architecture. 
The network is composed of two branches: 
one branch uses the lighting MLP to predict the {\em view-dependent} specular color $\textbf{c}_s$ and 
the other branch uses the palette MLP to output the {\em view-independent} diffuse color $\textbf{c}_d$. 

For the specular color branch, we use a structure similar to the Instant-NGP's color function. The lighting MLP also ingests geometry features output from the density MLP and {\em spherical harmonics} (SH) encoded view direction as input. The only difference is that the output of the lighting MLP is a grayscale color instead of the original RGB color. Before optimization, we leverage the hidden parameters of the base NeRF's color function to initialize the shallow layers of the palette MLP to improve the prediction of $\textbf{c}_s$ at the early stage.

\begin{figure}[htb]
\begin{center}
		$\begin{array}{c@{\hspace{0.01in}}c@{\hspace{0.01in}}c}
				\includegraphics[height=0.8in]{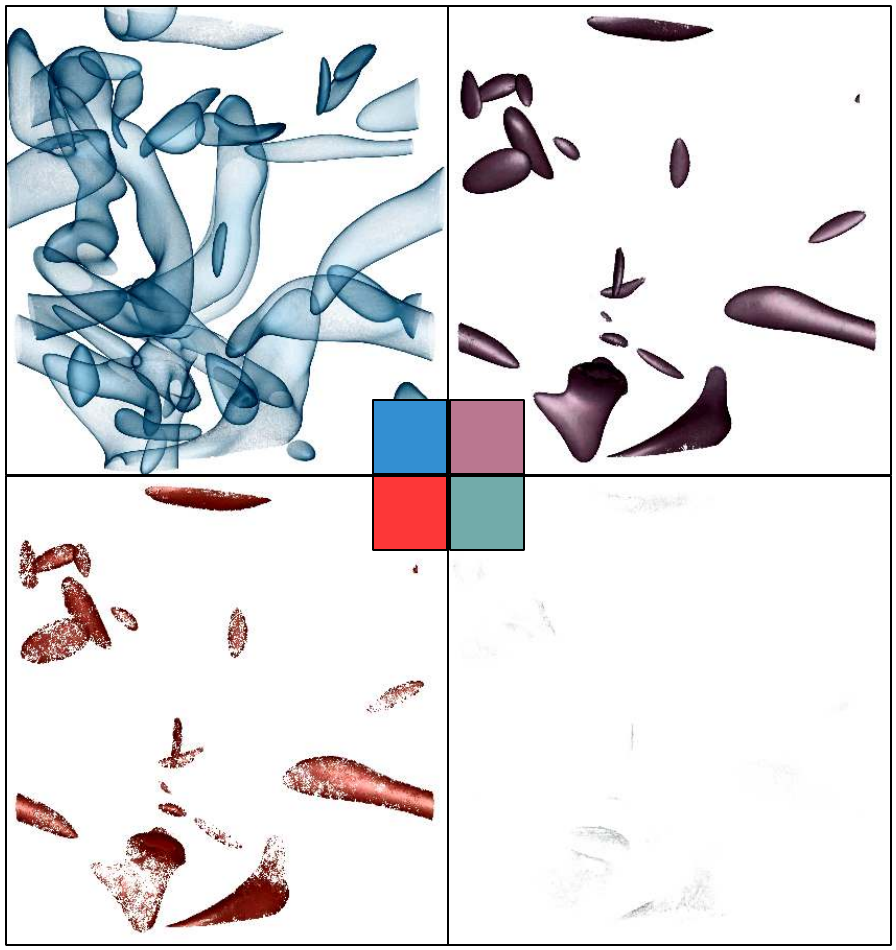} &
				\includegraphics[height=0.65in]{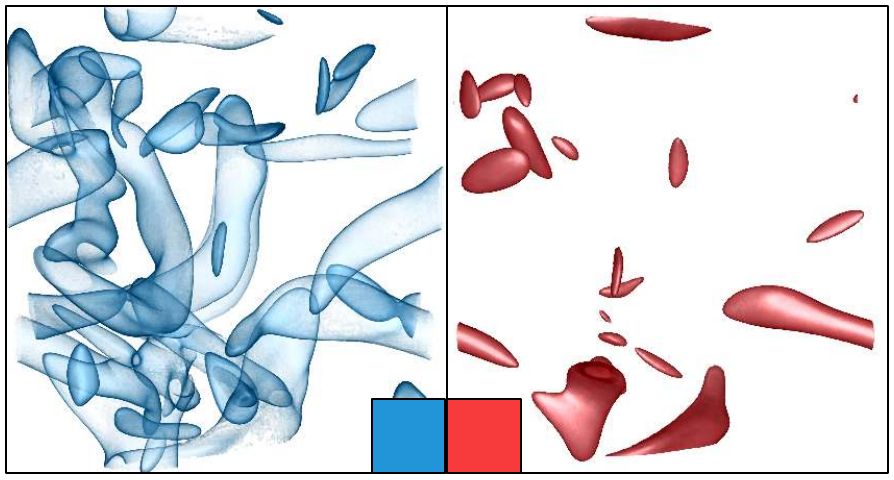} &
				\includegraphics[height=0.65in]{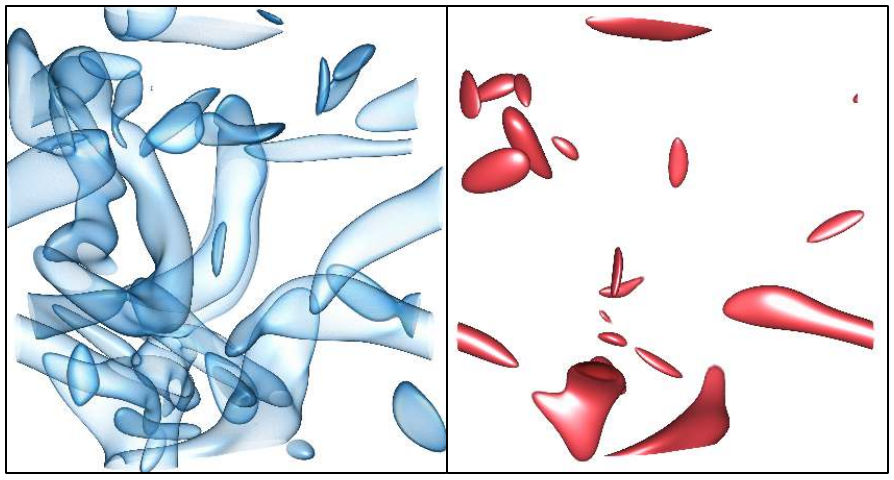}\\
				\includegraphics[height=1.0in]{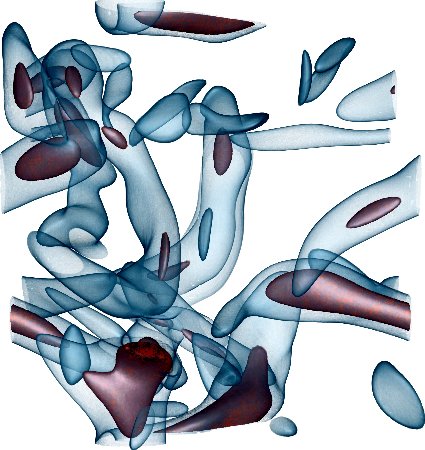} &
				\includegraphics[height=1.0in]{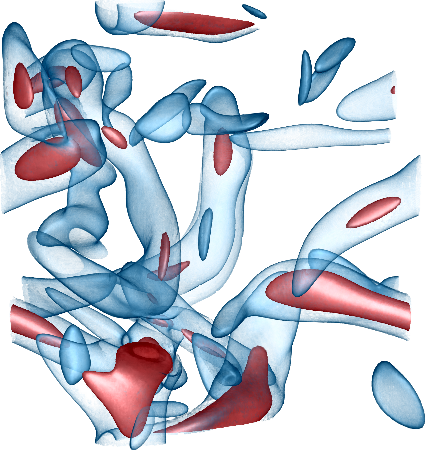} &
				\includegraphics[height=1.0in]{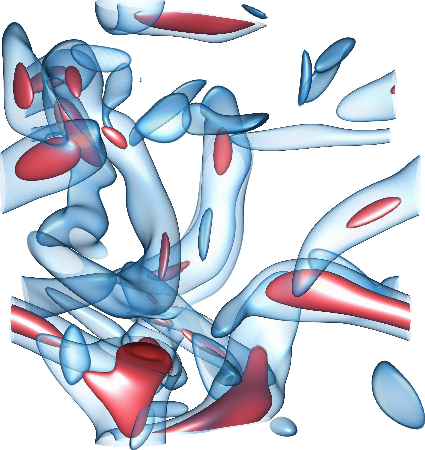}  \\
				\mbox{\footnotesize (a) w/o refinement}
				& \mbox{\footnotesize (b) w/ refinement}
				& \mbox{\footnotesize (c) GT}
			\end{array}$
	\end{center}
	\vspace{-.25in}
\caption{Classification and rendering results without and with palette refinement. Top: classification results for each palette color. Bottom: rendering results when considering all palette colors.} 
\label{fig:palette-refine}
\end{figure}

For the diffuse color branch, we design a palette MLP that captures the weights $\omega$ of palette colors $\textbf{P}$ for any points within the RF.
This way, we can use $\omega$ to classify points in the RF according to their primary colors, preserving visual content consistency across different views during style editing.
Similar to the recent NeRF's palette-based recoloring strategy~\cite{Gong-RecolorNeRF, Kuang-PaletteNeRF, Tojo-CGF22}, we apply the RGB convex hull method~\cite{Tan-TOG18} to extract colors $\textbf{P}_{\convex}$ from training images. However, unlike previous palette recoloring methods that directly employ $\textbf{P}_{\convex}$ to initialize the palette, we further refine $\textbf{P}_{\convex}$ as it usually contains many similar colors that may significantly hinder the classification. We remove similar colors when the distance between two colors is below a certain threshold. This distance is measured by the {\em hue} component of colors in the hue-saturation-brightness (HSB) space. The {\em saturation} and {\em brightness} components of palette colors are not considered in measuring the distance as they can be easily optimized during training. Figure~\ref{fig:palette-refine} gives an example that compares classification and rendering results without and with palette refinement. Refer to the appendix for more details about the refinement algorithm.

Once the palette colors are initialized and the number of palette colors $N_P$ is determined, we construct and optimize the PCN. During the optimization, the density MLP and hash-grid encoder $\textbf{E}_{\sigma}$ learned in the first stage are fixed for invariant density representation. The palette MLP takes features output from the palette hash-grid encoder $\textbf{E}_p$ as input and outputs three values: 
an intensity value $I$ shared by all palette colors to counteract the range shift caused by the RGB convex hull color normalization, 
a set of weights $\omega$ that indicate the contribution of each palette color to $\textbf{c}_d$, and 
a color offset vector $\delta$ to enhance the expressiveness of palette colors. 
Given the refined palette colors $\textbf{P}$ and specular color $\textbf{c}_s$, the output color $\textbf{c}$ associated with position $\textbf{x}$ and view direction $\textbf{d}$ is computed as
\begin{equation}
\label{eqn:paletteColor}
	\textbf{c}(\textbf{x},\textbf{d}) = \textbf{c}_s(\textbf{x},\textbf{d})+I(\textbf{x})\sum_{i=1}^{N_P}\omega_i(\textbf{x})(P_i+\delta_i(\textbf{x})),
\end{equation}
where $\textbf{I}(\textbf{x})$ is the intensity value at position $\textbf{x}$, $\omega_i(\textbf{x})$ and $\delta_i(\textbf{x})$ denote the weight and color offset for palette color $P_i$ at $\textbf{x}$, respectively.
We then predict pixel color $\hat{\textbf{c}}(\textbf{r})$ using Equation~\ref{eqn:baseNeRFColor} for $\textbf{c}(\textbf{x},\textbf{d})$, and optimize network parameters and palette colors with loss $\mathcal{L}_{\palette}$ defined as
\begin{equation}
\label{eqn:paletteLoss}
	\mathcal{L}_{\palette}=||\textbf{c}(\textbf{r}) - \hat{\textbf{c}}(\textbf{r})||^{2}_2 + \lambda_{\delta}\sum_{i=1}^{M}\sum_{j=1}^{N_P}||\delta_j(\textbf{x}_i)||^{2}_2.
\end{equation}
The first term of $\mathcal{L}_{\palette}$ is the MSE loss between the GT pixel color $\textbf{c}(\textbf{r})$ and predicted pixel color $\hat{\textbf{c}}(\textbf{r})$. 
The second term is the offset regularization loss for sample points along ray $\textbf{r}$, aiming to suppress the magnitudes of color offsets and avoid significant palette color shiftings. We set $\lambda_{\delta} = 0.1$ to control the regularization strength.









\vspace{-0.05in}
\subsection{Photorealistic Style Editing (PSE)}
\label{subsec:PSE}

When inferring the PCN, the output palette weights $\omega(\textbf{x})$ can classify any 3D position within the RF according to its palette color. We then tune the network values to support PSE. Interactive PSE is achieved using Instant-NGP's fast inference ability.

We use the HSB space for recoloring following~\cite{Kuang-PaletteNeRF}. Given the target palette colors $\textbf{P}'$, we compute the difference $\Delta \textbf{P}$ between the original palette colors $\textbf{P}$ and $\textbf{P}'$ in the HSB space. $\Delta P_i$ are then added 
to $P_i+\delta_i(\textbf{x})$
in Equation~\ref{eqn:paletteColor}. 
To achieve opacity or lighting editing, we multiply $\sigma(\textbf{x}_{P_i})$ or $\textbf{c}_s(\textbf{x}_{P_i}, \textbf{d})$ by a scalar value, where $\textbf{x}_{P_i}$ denotes a point position receiving its primary color contribution from $P_i$. 

%
%

\vspace{-0.05in}
\subsection{Unrestricted Color Network (UCN)}
\label{subsec:UCN}
Although the PCN can represent the colors of the RF with PSE, it prohibits using diverse colors in NPSE. 
To address this issue, we utilize KD to optimize a UCN (student model) that produces similar diffuse colors within the PCN (teacher model) with no color palette constraint. 

Before distillation, we initially extract the average palette colors $\bar{\textbf{P}}$ from each reference style to replace the original palette colors $\textbf{P}$.
This color transfer step ensures the distilled student network can match with styles, reducing optimization effort in NPSE. 
Our UCN comprises one hash-grid encoder $\textbf{E}_u$ following an unrestricted MLP that outputs color $\textbf{c}_u$. 
During distillation, we first randomly sample camera views from the scene. Then we optimize $\textbf{E}_u$ with volume-aligned loss $\mathcal{L}_{\volume}$ adopted from~\cite{Fang-AAAI23} for each view. $\mathcal{L}_{\volume}$ is defined as
\begin{equation}
	\mathcal{L}_{\volume} = ||\textbf{E}_p(\textbf{x}) - \textbf{E}_u(\textbf{x})||^{2}_2,
\end{equation}
where $\textbf{E}_p(\textbf{x})$ and $\textbf{E}_u(\textbf{x})$ are the output features from the palette and unrestricted encoders for each sample point $\textbf{x}$. We optimize $\mathcal{L}_{\volume}$ for 150 iterations to initialize $\textbf{E}_u(\textbf{x})$ to accelerate the subsequent distillation. For the following 500 iterations, we optimize both $\textbf{E}_u(\textbf{x})$ and unrestricted MLP with color loss $\mathcal{L}_{\colors}$, which is defined as
\begin{equation}
	\mathcal{L}_{\colors} = ||\bar{\textbf{c}}_d(\textbf{x}) - \textbf{c}_u(\textbf{x})||^{2}_2,
\end{equation}
where $\bar{\textbf{c}}_d(\textbf{x})$ and $\textbf{c}_u(\textbf{x})$ are the diffuse colors output from the PCN with the average palette colors $\bar{\textbf{P}}$ and the UCN at $\textbf{x}$.  

\vspace{-0.05in}
\subsection{Non-Photorealistic Style Editing (NPSE)}
\label{subsec:NPSE}

Given one or multiple reference images, we first extract $N_P$ desired reference styles $\textbf{S}=\{S_1, S_2, \cdots, S_{N_P}\}$ for $N_P$ regions of the RF. 
When the number of reference images is less than $N_P$, or users are only interested in local patterns of one image, 
we can leverage unsupervised segmentation techniques such as the segment anything model (SAM)~\cite{Kirillov-arxiv23} to automatically or manually (with point prompt selection) divide a reference image into localized style regions for flexible stylization.

\begin{figure}[htb]
\begin{center}
\includegraphics[width=0.85\linewidth]{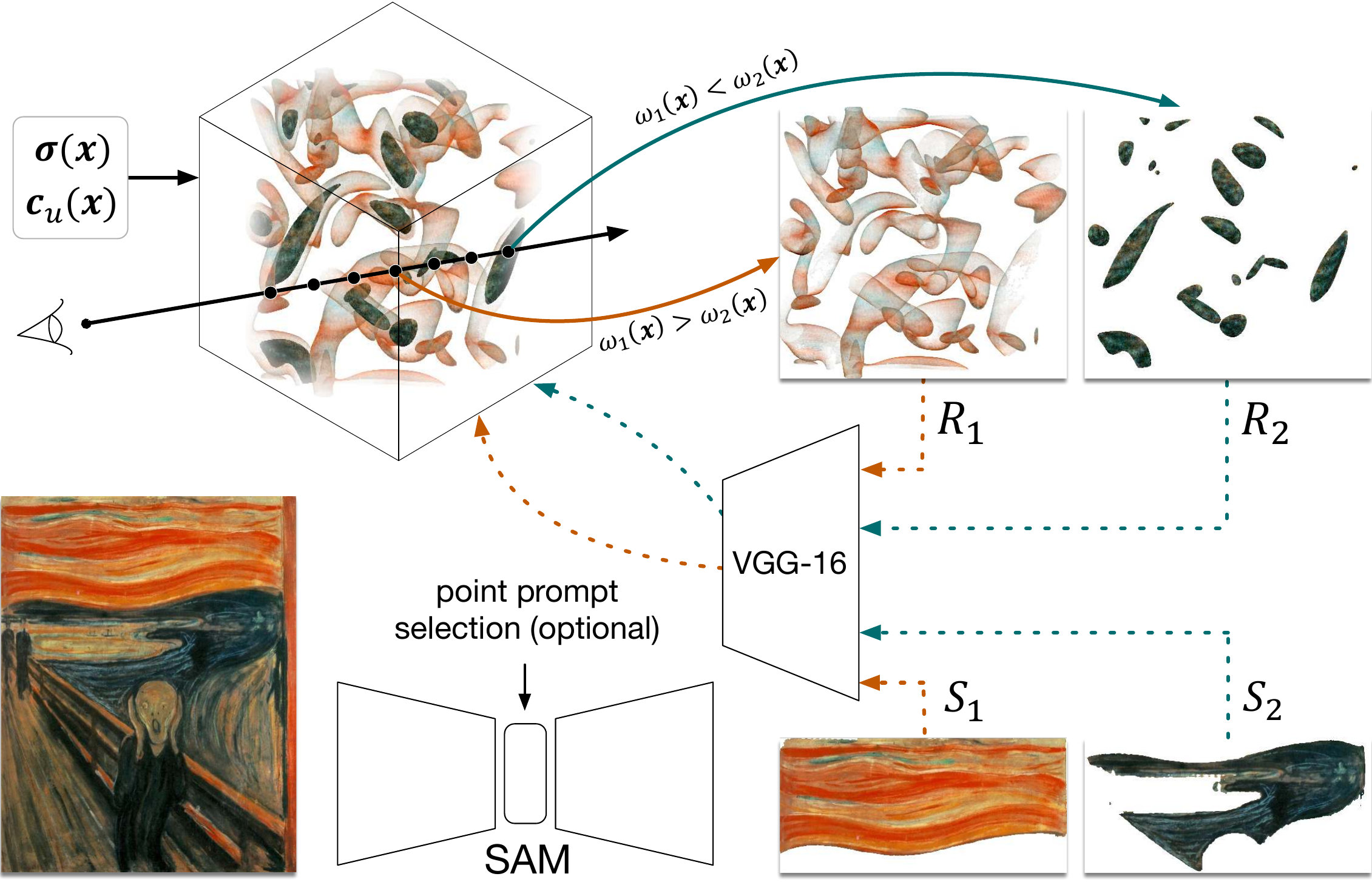}
\end{center}
\vspace{-.25in}
\caption{The process of NPSE. We start segmenting one reference image into styles $(S_1, S_2)$ via SAM. 
Each rendering $(R_1, R_2)$ showing separate content is then stylized by one assigned style.} 
\label{fig:NPR-pipeline}
\end{figure}

Figure~\ref{fig:NPR-pipeline} shows our training process for non-photorealistic style optimization. We use the optimizable unrestricted color $\textbf{c}_u(\textbf{x})$ (Section~\ref{subsec:UCN}) and frozen density $\sigma(\textbf{x})$ of the base NeRF (Section~\ref{subsec:base-nerf}) to construct a new RF. 
During training, we first compute $N_P$ rendering images $\textbf{R}=\{R_1, R_2, \cdots, R_{N_P}\}$ from the new RF for each camera view. 
For one sample point $\textbf{x}$, it only contributes to $R_i$ if its palette weight $\omega_i(\textbf{x})$ is greater than other palette weights.
Once we have $N_P$ renderings $\textbf{R}$ and styles $\textbf{S}$, a pre-trained VGG-16 model~\cite{Simonyan-ICLR15} computes a stylization loss based on a user-specified or randomly assigned style mapping. Specifically, we use the {\em nearest neighbor feature matching} (NNFM) loss~\cite{Zhang-ARF, Kolkin-NNST} for each rendering and style, which is formulated as
\begin{equation}
	\mathcal{L}_{\nnfm}(\textbf{R},\textbf{S})=\frac{1}{(N_FN_P)}\sum_{i=1}^{N_P}\sum_{f_j\in\phi(R_i)}\min_{f_k\in\phi(S_i)}d(f_j,f_k),
\end{equation}
where $\phi(R_i)$ and $\phi(S_i)$ are the rendering and style feature vectors extracted from the VGG-16 model $\phi$. $N_F$ is the number of feature vectors for $\phi(R_i)$, and $d$ is the cosine distance between feature vectors $f_j$ and $f_k$.

\begin{figure}[htb]
\begin{center}
		$\begin{array}{c@{\hspace{0.05in}}c@{\hspace{0.05in}}c}
				\includegraphics[width=0.3\linewidth]{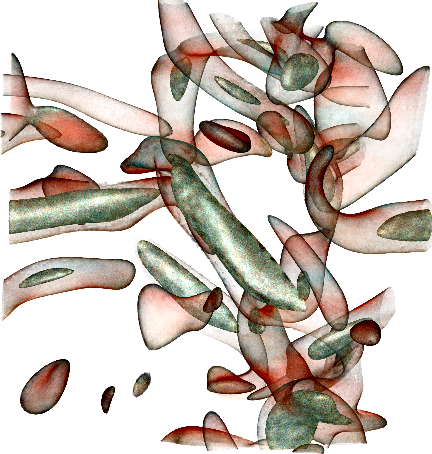} &
				\includegraphics[width=0.3\linewidth]{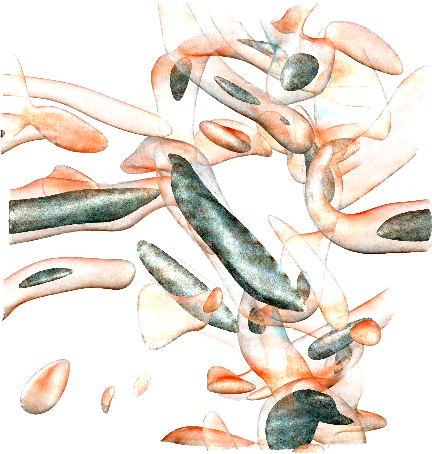} &
				\includegraphics[width=0.3\linewidth]{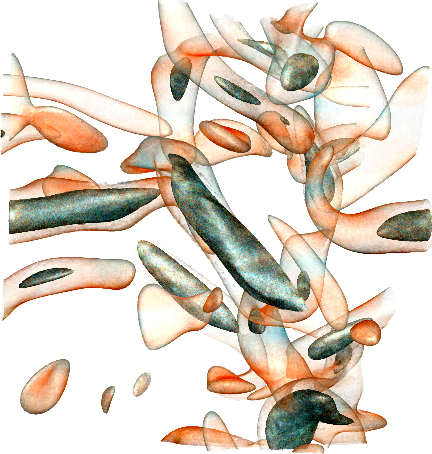} \\
				\mbox{\footnotesize (a) black}
				& \mbox{\footnotesize (b) white}
				& \mbox{\footnotesize (c) luminance}
			\end{array}$
	\end{center}
	\vspace{-.25in}
\caption{Inference results using different background colors to optimize the NNFM loss. (c) yields the best stylization outcome.} 
\label{fig:bg-color}
\end{figure}

Unlike previous NeRF-based non-photorealistic stylization methods~\cite{Zhang-ARF, Huang-CVPR22, Liu-CVPR23}, our style transfer process involves three unique designs for the VolVis scenario. 
First, different from traditional NeRF-based stylization that directly discards the view-dependent color to maintain visual content consistency across the views, we optimize $\textbf{c}_u(\textbf{x})$ (i.e., diffuse color) during training and output $\textbf{c}_u(\textbf{x})$+$\textbf{c}_s(\textbf{x},\textbf{d})$ (i.e., diffuse and specular colors) during inference. The rationale behind this design is to prevent the view-dependent color from influencing visual content consistency and preserve the specular lighting effect in the final output. 
Second, when rendering $\textbf{R}$ in training, we observe that varying background colors could impact the stylization results (refer to an example shown in Figure~\ref{fig:bg-color}). This phenomenon is caused by the inherent transparency in the DVR scene, which is not often present in natural scenes. Consequently, different background colors could lead to distinct renderings using the identical network, potentially incurring unsatisfactory stylization. 
To address this issue, we extract the luminance value from each style $S_i$ as the corresponding background color for each rendering $R_i$.
Third, we omit the content loss between the original and stylized images due to suboptimal stylization outcomes.

\begin{figure}[htb]
\begin{center}
\includegraphics[width=0.95\linewidth]{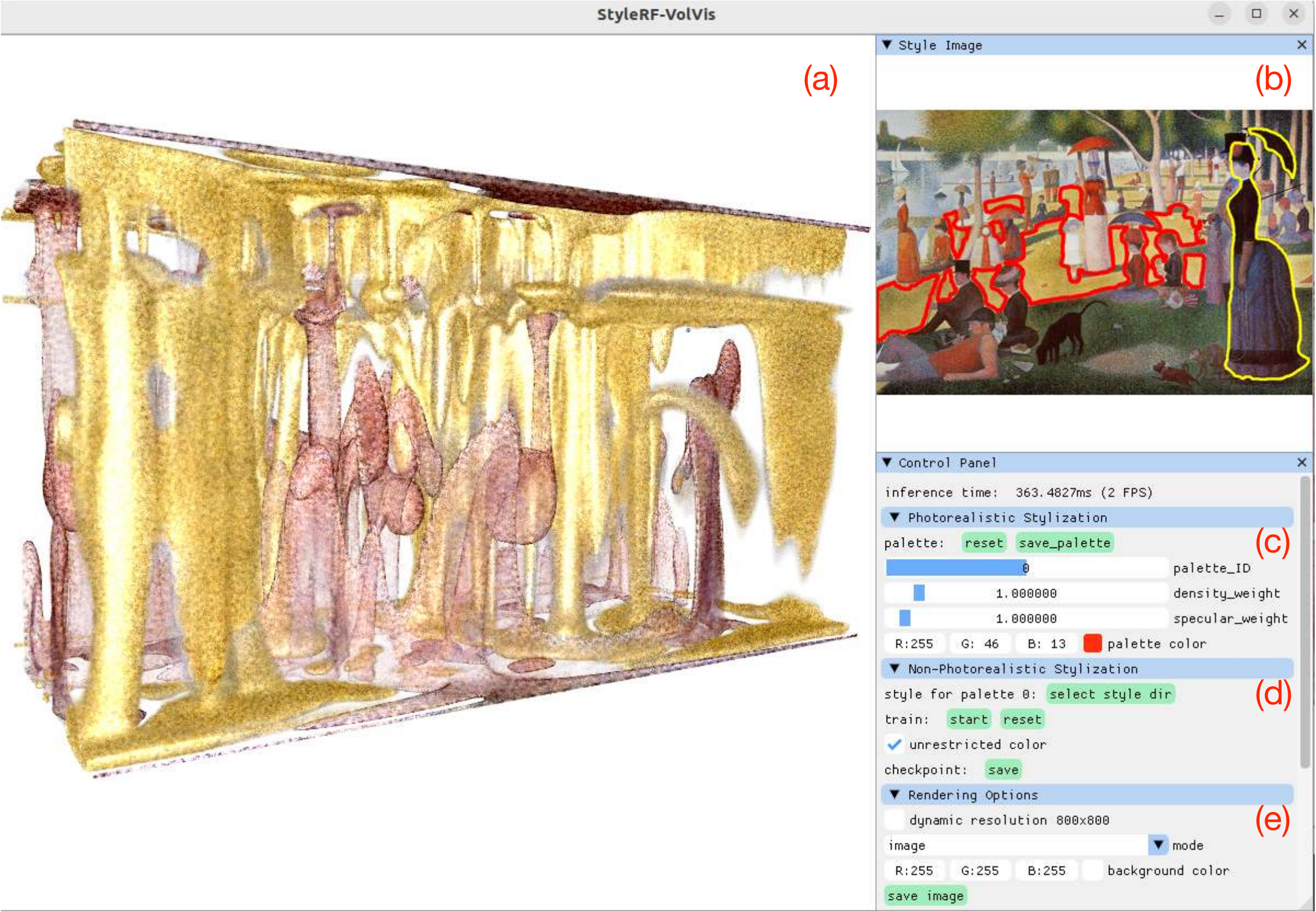}
\end{center}
\vspace{-.225in}
\caption{The screenshot of StyleRF-VolVis interface. (a) the stylization result of the DVR scene; (b) the reference image with selected styles highlighted with segmentation boundaries; (c) the PSE panel for color, density, and lighting editing; (d) the NPSE panel for style selection and UCN parameter training, resetting, and saving; (e) rendering options for adjusting the background color or saving stylized images, etc.} 
\label{fig:GUI}
\end{figure}

\vspace{-0.05in}
\subsection{Interactive interface}
\label{subsec:GUI}
Figure~\ref{fig:GUI} shows our visual interface for users to conduct PSE and NPSE for a given DVR scene. 
For PSE, users select a palette ID and interactively modify the corresponding scene region's color, opacity, and lighting. 
For NPSE, users first specify the target style for each scene region from single or multiple reference images. 
They then start optimizating UCN. 
After NPSE, users can still change the opacity and lighting of the stylized DVR scene but cannot modify the color as the stylized scene is represented with UCN. 
Refer to the accompanying video for the recorded interaction with the interface.

\begin{figure*}[!htb]
\begin{center}
	$\begin{array}{c}
		\includegraphics[width=0.95\linewidth]{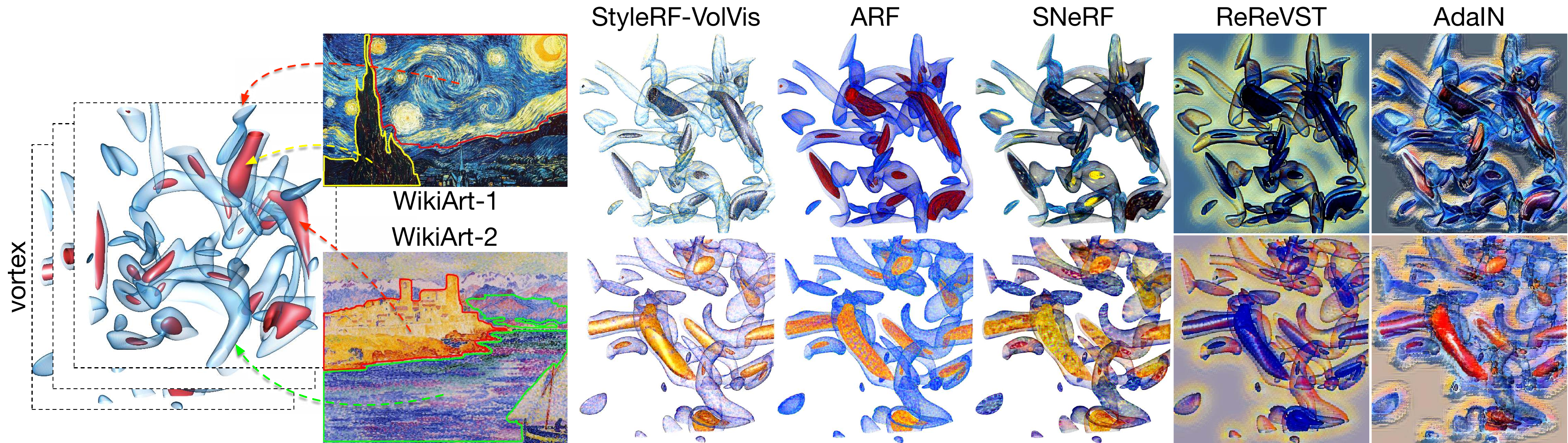} \\
		\includegraphics[width=0.95\linewidth]{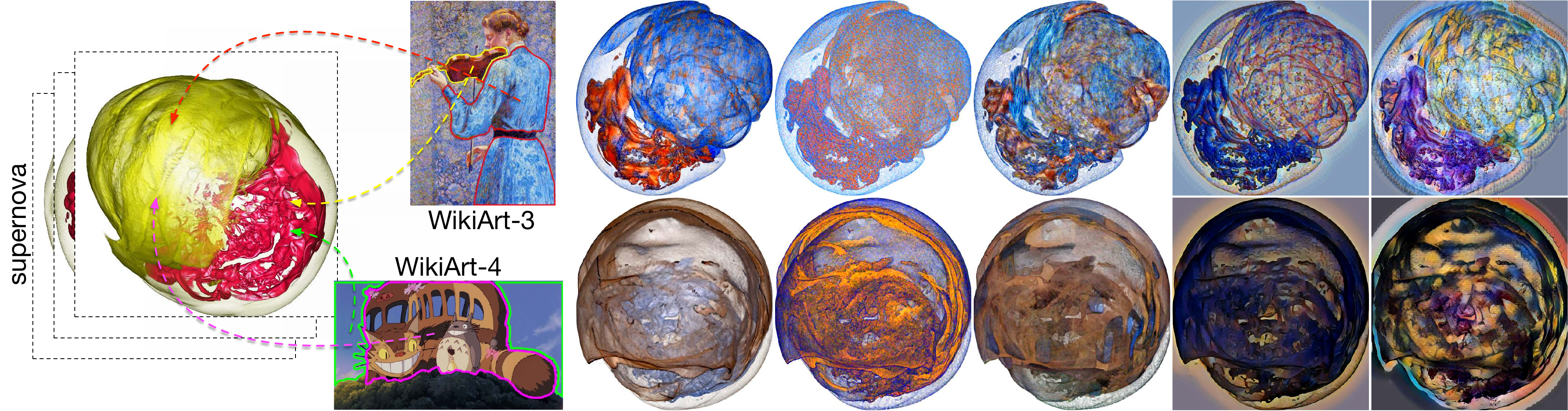} \\
		\includegraphics[width=0.95\linewidth]{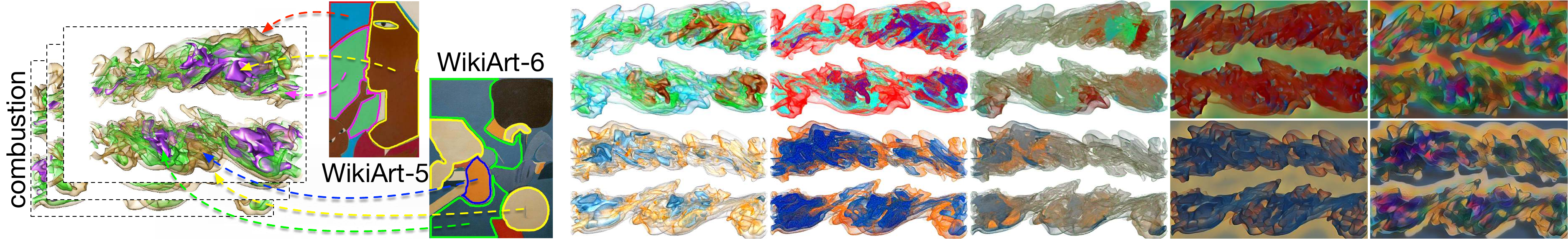} \\
		\includegraphics[width=0.95\linewidth]{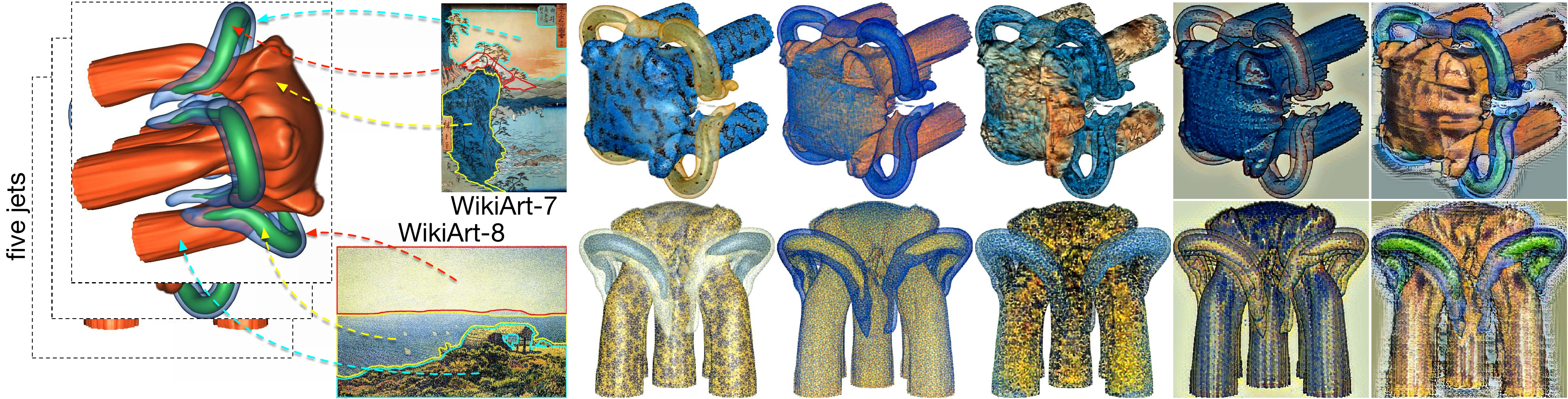} \\
	\end{array}$
	\end{center}
	\vspace{-.275in}
\caption{Comparison of VolVis stylization. We provide two reference images for each dataset. For StyleRF-VolVis, we highlight extracted style regions and their correspondence with the DVR scene regions. None of the other methods support customized style selection and content mapping.} 
\label{fig:baseline}
\end{figure*}

\begin{table}[htb]
\caption{The datasets and their respective settings.}
\vspace{-0.1in}
\centering
\resizebox{\columnwidth}{!}{
\begin{tabular}{c|ccccc}
   	    & volume &\# visible  &\# training  & \# inference  & image     \\ 
 dataset & resolution &  ranges  & images &  images & resolution    \\ \hline
 \pin{aneurysm} & \pin{256$\times$256$\times$256}  &\pin{1} &\pin{92} & \pin{181}  &\pin{$800\times800$} \\
 combustion &  480$\times$720$\times$120 &3 &92 & 181  &$800\times800$  \\
 earthquake & 256$\times$256$\times$96  &3 &92 & 181  &$800\times800$ \\
 five jets & 128$\times$128$\times$128  &3 &92 & 181  &$800\times800$ \\
 mantle & 360$\times$201$\times$180  &2 &92 & 181 &$800\times800$ \\
 \pin{rotstrat} & \pin{4096$\times$4096$\times$4096}  &\pin{2} &\pin{92} & \pin{181}  &\pin{$800\times800$} \\
 solar plume & 128$\times$128$\times$512 &2 &92 & 181  &$800\times800$ \\
 supernova & 432$\times$432$\times$432 &2 &92 & 181  &$800\times800$ \\
 vortex   & 128$\times$128$\times$128  &2 &92 & 181  &$800\times800$ \\
\end{tabular}
}
\label{tab:dataset}
\end{table}

\vspace{-0.1in}
\section{Results and Discussion}

\subsection{Datasets, Training, Baselines, and Metrics}

\textbf{Datasets and network training.}
To show the quality, consistency, and flexibility of StyleRF-VolVis, we evaluate it using the datasets listed in Table~\ref{tab:dataset}. 
The visible ranges are the opacity bumps in the TF specification corresponding to distinct visual contents for generating the original rendering. 
All reference images used are copyright-free, coming from WikiArt~\cite{WikiArt, WikiArt-old} and Pexels~\cite{Pexels}. 
We implemented StyleRF-VolVis using PyTorch and ran experiments on an NVIDIA RTX A4000 graphics card with 16 GB memory. 
We trained the base NeRF model with 30,000 iterations and PCN with 2,500 iterations using 92 DVR images with cameras evenly placed along a sphere enclosing the volume. 
The KD of UCN and NPSE does not need GT images. 
For KD, we optimized UCN with 624 iterations, where each iteration trains on one of the randomly sampled 312 camera views. 
For NPSE, we optimized UCN using the NNFM loss with 210 iterations, where each iteration trains on one of the uniformly sampled 42 camera views. 
For all three stages (base NeRF, PCN, and UCN), we applied the Adam optimizer with a learning rate of 0.01 and $\beta_1=0.9$, $\beta_2=0.999$ for training and set the batch size to 4,096 rays.
For the base NeRF model, PCN, and UCN during KD, we decayed the learning rate exponentially for every iteration until it reached 0. 
During NPSE, we fixed the learning rate and followed the setting of ARF~\cite{Zhang-ARF}. 
We used the $conv3$ block of VGG-16 to extract features for computing the NNFM loss. 
{\em Deferred backpropagation} was employed to update UCN's parameters for the entire image under limited GPU memory. 
During inference, we generated 181 stylized images with novel views, with cameras evenly placed along a trajectory to capture the full 360-degree view. 

\textbf{Baselines.}
We compare four baseline methods for NPSE:
\begin{myitemize}
\vspace{-0.05in}
	\item ARF~\cite{Zhang-ARF} is a NeRF-based method that trains a NeRF model to represent the scene and then updates the pre-trained NeRF parameters with NNFM loss to generate stylized novel view images. 
	\item SNeRF~\cite{Nguyen-Phuoc-SNeRF} is a NeRF-based method that first constructs the scene with a NeRF model and then alternatively finetunes the parameters of the pre-trained NeRF and an additional image stylization module to achieve stylization. 
	\item ReReVST~\cite{Wang-ReReVST} is a video-based stylization method. It first performs novel view synthesis with a NeRF model, then arranges the novel view images as a video sequence, and finally performs video stylization to yield the result. 
	\item AdaIN~\cite{Gatys-CVPR16} is an image-based stylization method that optimizes a NeRF model for generating novel view images of the scene and then applies stylization to the NeRF synthesized images.
\vspace{-0.05in}
\end{myitemize}
\noindent Note that we did not compare StyleRF-VolVis with the more recent StyleRF~\cite{Liu-CVPR23} work as it mainly focuses on the ability of zero-shot transfer to unseen reference images. 
StyleRF trains on all reference images from the WikiArt dataset to ensure better generalization for unseen reference images. However, this could lead to inferior stylization results compared to optimization methods like ARF and SNeRF.  
Existing visualization generation methods in scientific visualization~\cite{Berger-TVCG19, Hong-DNN-VolVis, He-InsituNet, Han-TVCG23} were also excluded because they were not designed for style transfer. 

\textbf{Evalutation metrics.}
Since 3D style transfer is relatively novel and scarcely explored, limited metrics are available for quantitatively assessing stylization quality. 
Consequently, our evaluation focuses on {\em cross-view consistency} of stylization results. 
Specifically, we generated testing videos in which each frame represents a rendered image from one of the 181 novel viewpoints of our stylized scene. Then, we warped one view to the other using {\em softmax splatting}~\cite{Simon-CVPR20} according to the optical flow estimated with RAFT~\cite{Zachary-ECCV20}. Finally, we computed the MSE and LPIPS~\cite{Zhang-CVPR18} scores to measure the cross-view consistency. 
Similar to~\cite{Wang-ReReVST, Nguyen-Phuoc-SNeRF, Liu-CVPR23}, we calculated the {\em long-range} consistency for faraway views (with an interval of 10 among the 181 inference views) and {\em short-range} consistency for adjacent views, respectively.

\vspace{-0.05in}
\subsection{Qualitative Comparison}

We compare StyleRF-VolVis against baseline methods using DVR scenes of vortex, supernova, combustion, and five jets. Each scene showcases two stylization results using distinct reference images. Figure~\ref{fig:baseline} shows the qualitative comparison. Unlike all baseline methods, StyleRF-VolVis supports customized style selection and mapping to distinct visual contents, which we highlight in the DVR scene.

Overall, StyleRF-VolVis produces clearer and more consistent stylization results than other methods. 
In contrast to video- and image-based stylization methods (ReReVST and AdaIN), NeRF-based stylization methods (StyleRF-VolVis, ARF, and SNeRF) preserve better geometry consistency and are capable of separating foreground objects from the background during stylization. 
Among NeRF-based methods, SNeRF captures the overall style of the reference image but cannot assign distinct styles to different visual regions of the original scene. Thus, SNeRF's stylization is blurry and mixed, making it hard to identify the difference between individual visual contents in the original DVR scene.
For the vortex dataset, ARF shows better stylization than SNeRF in preserving visual content distinction. 
However, the uncontrollable ARF style selection process could let the model choose a negligible local style that does not match the overall style (e.g., the WikiArt-1 case) or select one style for multiple visual regions (e.g., the WikiArt-8 case). Both drawbacks could lead to poor stylization results.
In addition, in the stylization process, ARF and SNeRF discard view directions in their input to ensure cross-view stylization consistency. Consequently, they cannot preserve the view-dependent lighting of the DVR scene.
Unlike ARF and SNeRF, StyleRF-VolVis utilizes the lighting MLP to preserve lighting information of visual content while ensuring cross-view consistency during NPSE by removing view direction input in UCN optimization. 
Moreover, StyleRF-VolVis allows explicit assigning of different styles to different DVR regions. Via color transfer, we ensure that the region color always aligns with the overall style color, avoiding the disadvantage of NNFM loss, which focuses on negligible local style. 
\begin{table}[htb]
\caption{Comparing the total training time and per-image inference time for NeRF-based methods. The best ones are shown in bold.}
\vspace{-0.1in}
\centering
{\scriptsize
\begin{tabular}{c|c|rr}
dataset & method  &training time &inference time \\ \hline
      		&ARF    &\textbf{14.6 m} &204 ms \\
vortex 	&SNeRF &3.6 h  &\textbf{173 ms}	 \\
 			&StyleRF-VolVis  &29.7 m &373 ms \\ \hline
  		&ARF	&\textbf{20.1 m}   &210 ms	 \\
combustion  &SNeRF &3.8 h  &\textbf{195 ms} \\
			&StyleRF-VolVis  &1.2 h &564 ms  \\
\end{tabular}
}
\label{tab:comp-time}
\end{table}

\vspace{-0.05in}
\subsection{Quantitative Comparison}

{\bf Training and inference time.} 
We report the training and inference time for all NeRF-based methods in Table~\ref{tab:comp-time}.
The training of StyleRF-VolVis is slower than ARF but faster than SNeRF.
Unlike ARF, StyleRF-VolVis requires additional steps to optimize PCN and apply KD to UCN before NPSE. 
Moreover, optimizing on a single camera view during NPSE requires rendering different regions independently. 
This process maps a style to a region, slowing the stylization process. 
Although the alternative training process of SNeRF is relatively straightforward, it demands significant time to train the image stylization module, leading to the longest training time. 
For inference, StyleRF-VolVis needs additional feedforward steps of the lighting MLP to provide lighting information, which takes longer than ARF or SNeRF. 

%
\begin{table}[htb]
\caption{Average short- and long-range cross-view consistency for stylization cases shown in Figure~\ref{fig:baseline}.} 
\vspace{-0.1in}
\centering
{\scriptsize
\begin{tabular}{c|cccc}
  &\multicolumn{2}{c}{short-range} & \multicolumn{2}{c}{long-range}\\ 
 method		 & MSE$\downarrow$  & LPIPS$\downarrow$            & MSE$\downarrow$ &LPIPS$\downarrow$ \\\hline
AdaIN &0.087   &0.171   &0.118   &0.217   \\
ReReVST &0.049   &0.096   &\textbf{0.075}   &0.137   \\ \hline
ARF  &0.053   &0.056   &0.093   &0.105   \\
SNeRF &0.046   &0.060   &0.082   &0.106   \\
StyleRF-VolVis   &\textbf{0.045}   &\textbf{0.054}   &0.076   &\textbf{0.092}   \\
\end{tabular}
}
\label{tab:consistency}
\end{table}

{\bf Short- and long-range consistency.}
In Table~\ref{tab:consistency}, we compare short- and long-range cross-view consistency for different methods. 
We use MSE and LPIPS as the metrics. 
For each metric, we report the average value over eight stylization cases (WikiArt-1 to WikiArt-8) shown in Figure~\ref{fig:baseline}. 
StyleRF-VolVis achieves the best cross-view consistency, while AdaIN performs the worst as this image-based stylization method treats each image independently. 
ReReVST leads to comparable or even better MSEs than the NeRF-based methods because it produces stylization with relatively uniform colors, resulting in smaller pixel-wise errors. 
However, under the image-level LPIPS metric, the consistency of ReReVST significantly lags behind the NeRF-based methods.

\begin{table}[htb]
\caption{The votes of 14 participants on better solutions gathered from eight stylization cases (WikiArt-1 to WikiArt-8).}
\vspace{-0.1in}
\centering
\resizebox{\columnwidth}{!}{
\begin{tabular}{c|cccccccc|c}
 & \multicolumn{2}{c}{vortex} & \multicolumn{2}{c}{supernova} & \multicolumn{2}{c}{combustion}  & \multicolumn{2}{c|}{five jets}  &  \\ 
 method  & 1 & 2 & 3 & 4 & 5 & 6 & 7 & 8  & sum  \\ \hline
ARF  & 8 & 8 & 2 & 5 & 11 & 10 & 11 & \textbf{19} & 74 \\
SNeRF & \textbf{23} & 12 & \textbf{27} & 15 & 6 & 9 & 10 & 8 & 110 \\
StyleRF-VolVis   & 11 & \textbf{22} & 13 & \textbf{22} & \textbf{25} & \textbf{23}  & \textbf{21} & 15 & \textbf{152} \\
\end{tabular}
}
\label{tab:user-study}
\end{table}

{\bf Voting results from a user study.}
\pin{As the consistency metrics do not necessarily reflect the perceived style transfer quality, we conducted a user study to measure user preference for different stylizations following the University's IRB protocol.}
Per Figure~\ref{fig:baseline}, we included four datasets, each with two reference images, and considered three leading methods (ARF, SNeRF, and StyleRF-VolVis) for pairwise comparison. This leads to 24 (4$\times$2$\times$3) image pairs organized into eight (4$\times$2) groups. 
We recruited 14 students from a visualization class of undergraduate, Master's, and Ph.D.\ students in computer science and engineering, aerospace and mechanical engineering, and psychology majors. 

The participants were briefed on the evaluation criteria before proceeding to the study. 
A full-screen display shows each pair. At the top of each display, the original DVR and reference images (with extracted style regions highlighted in different color boundaries) are shown. At the bottom, the stylized images of two methods, randomly placed on the left and right sides and labeled `A' and `B,' are presented. 
The participants were asked to decide which one (`A' or `B') achieved the better stylization outcome, and no tie was allowed.
We asked the participants to take their time, as we did not record how much time they spent on each pair, each group, or the entire study. During the evaluation, they could go back and forth to update their votes. 
We advised them that many factors should be considered, including the overall impression, content preservation, style application, and visuals (color, opacity, and lighting). 
They could decide how to weigh them, and their criteria should be consistent throughout the study.
The entire study was completed in the classroom within 10 minutes. 

The voting results are shown in Table~\ref{tab:user-study}. 
We can see that StyleRF-VolVis wins for all stylization cases except WikiArt-1, WikiArt-3, and WikiArt-8. This may be attributed to the fact that some participants prefer 
the stylization result of ARF that provides a detailed local pattern for the WikiArt-8 case or 
the stylization results of SNeRF, which contain a more global pattern in WikiArt-1 (i.e., the dark tone and contrasting colors of {\em The Starry Night}) and WikiArt-3 cases. However, StyleRF-VolVis wins for all other cases and gains the most overall preference. The cross-view consistency of StyleRF-VolVis is also better than ARF and SNeRF, as shown in Table~\ref{tab:consistency}.

\begin{figure}[htb]
\begin{center}
\includegraphics[width=0.95\linewidth]{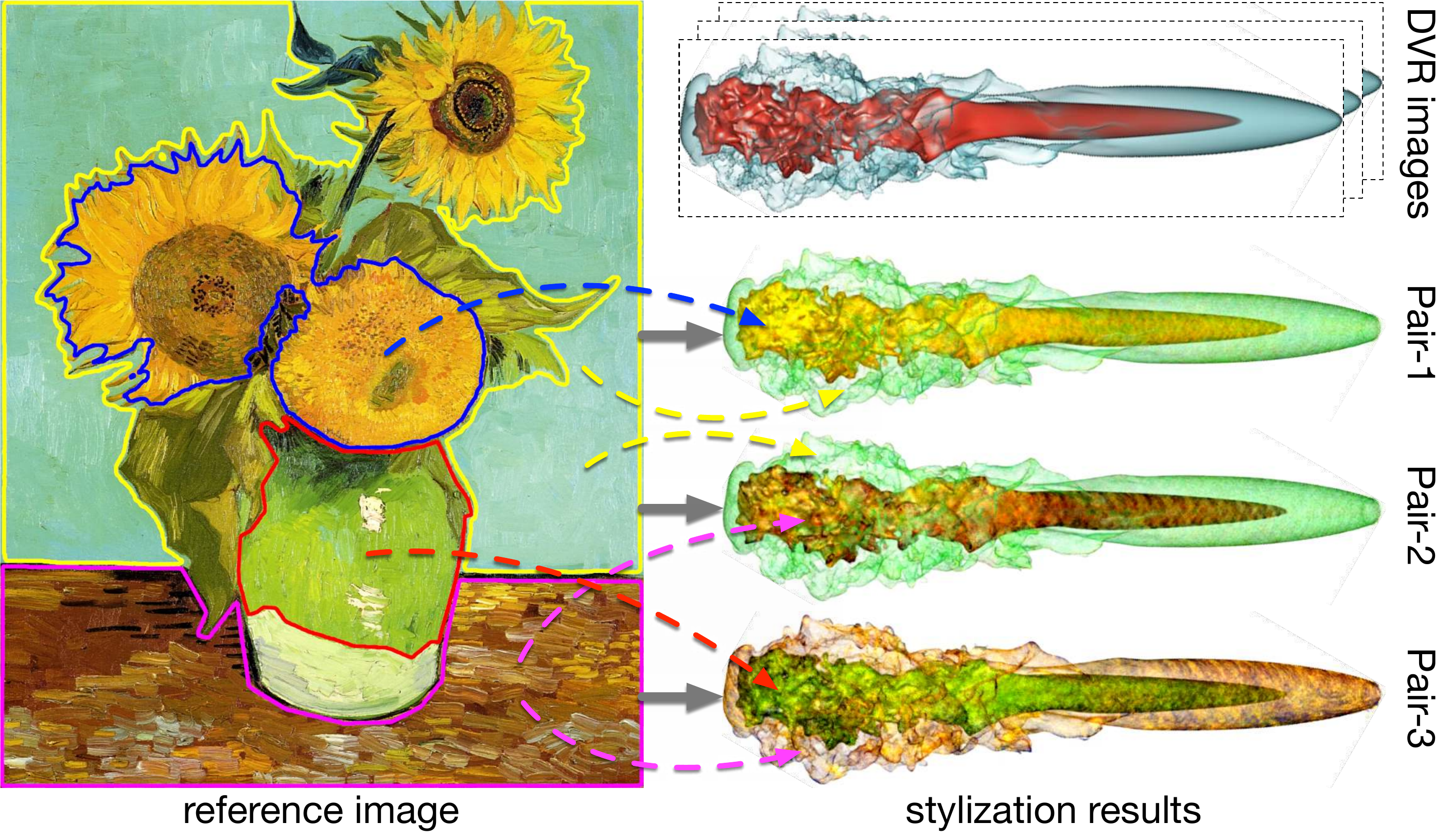}
\end{center}
\vspace{-.25in}
\caption{StyleRF-VolVis results on the solar plume dataset where three style pairs from one reference image are used for flexible stylization.}
\label{fig:OneImg2MultiStyles}
\end{figure}

\vspace{-0.05in}
\subsection{Flexibility of NPSE}
\label{sec:NPSE}

Our NPSE strategy allows users to define which style should be applied to which part of a DVR scene (analogous to using the TF), which provides more flexibility than other NPSE methods. Specifically, users can apply styles obtained from one reference image for multiple stylizations, leverage styles extracted from multiple reference images for one stylization, or only stylize one part of the DVR scene. All these additions are the unique features of StyleRF-VolVis.

{\bf One reference image for multiple stylizations.}
Unlike conventional NeRF-based stylization methods, which only produce one stylized scene from a single reference image, StyleRF-VolVis can generate various stylized outcomes with only one reference image.
Thanks to the advanced segmentation of PCN and SAM for the RF and reference image, we can extract multiple distinct styles and apply them to different scene regions.
By pairing styles and regions in various combinations, StyleRF-VolVis achieves controllable and diverse stylization outcomes.
Figure~\ref{fig:OneImg2MultiStyles} shows multiple stylization results of the solar plume DVR scene using the same reference image.
By comparing Pair-1 and Pair-2, we can see that when the style of one region remains unchanged, altering the style of another region keeps the unchanged style region intact. 
Further comparison of Pair-2 and Pair-3 reveals that applying the same style to different regions consistently transfers the expected texture.
These favorable features allow StyleRF-VolVis to offer users flexible stylization.
 
\begin{figure}[htb]
\begin{center}
\includegraphics[width=0.9\linewidth]{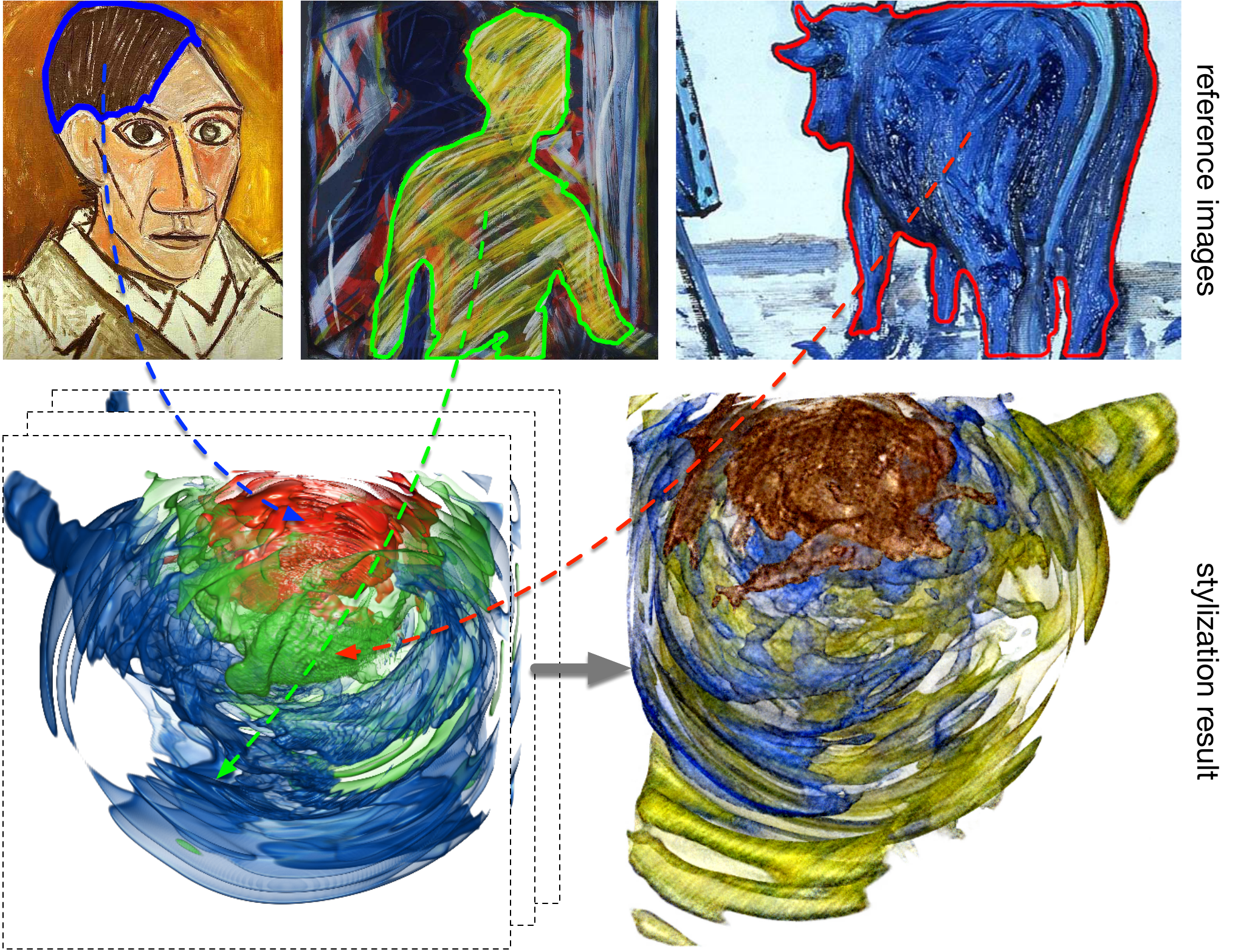}
\end{center}
\vspace{-.25in}
\caption{StyleRF-VolVis result on the earthquake dataset where three styles from different reference images are used for flexible stylization.} 
\label{fig:MultiStyles2OneImg}
\end{figure}

{\bf Multiple reference images for one stylization.}
Previous approaches like StyleRF~\cite{Liu-CVPR23} suggest leveraging extra supervision from large models pre-trained on extensive image datasets for segmenting scene elements to support style transfer from multiple reference images to a single scene. 
However, such a strategy cannot be directly applied to VolVis scenes. 
One reason lies in the inherent difference between DVR and natural scenes. 
DVR images exhibit more complex geometric relationships, with different visual contents often nested in layers with varied transparencies, rendering 2D segmentation methods futile. 
Another reason is that large models are typically pre-trained on natural images. 
Adapting such models to process DVR images would require extensive DVR images and substantial hardware resources for fine-tuning, a demand that is impractical for user-oriented VolVis scene stylization.
Our model can segment various regions via PCN, even though separate GT rendering results for different regions are missing during optimization and additional supervision from a 2D segmentation method is lacking. 
As illustrated in Figure~\ref{fig:MultiStyles2OneImg}, StyleRF-VolVis achieves stylized results for different regions using styles from multiple reference images. 
With a collection of reference images, StyleRF-VolVis could create vast combinations of different stylization results for one DVR scene.

\begin{figure}[htb]
\begin{center}
$\begin{array}{c@{\hspace{0.01in}}c@{\hspace{0.01in}}c}
		\includegraphics[height=1.2in]{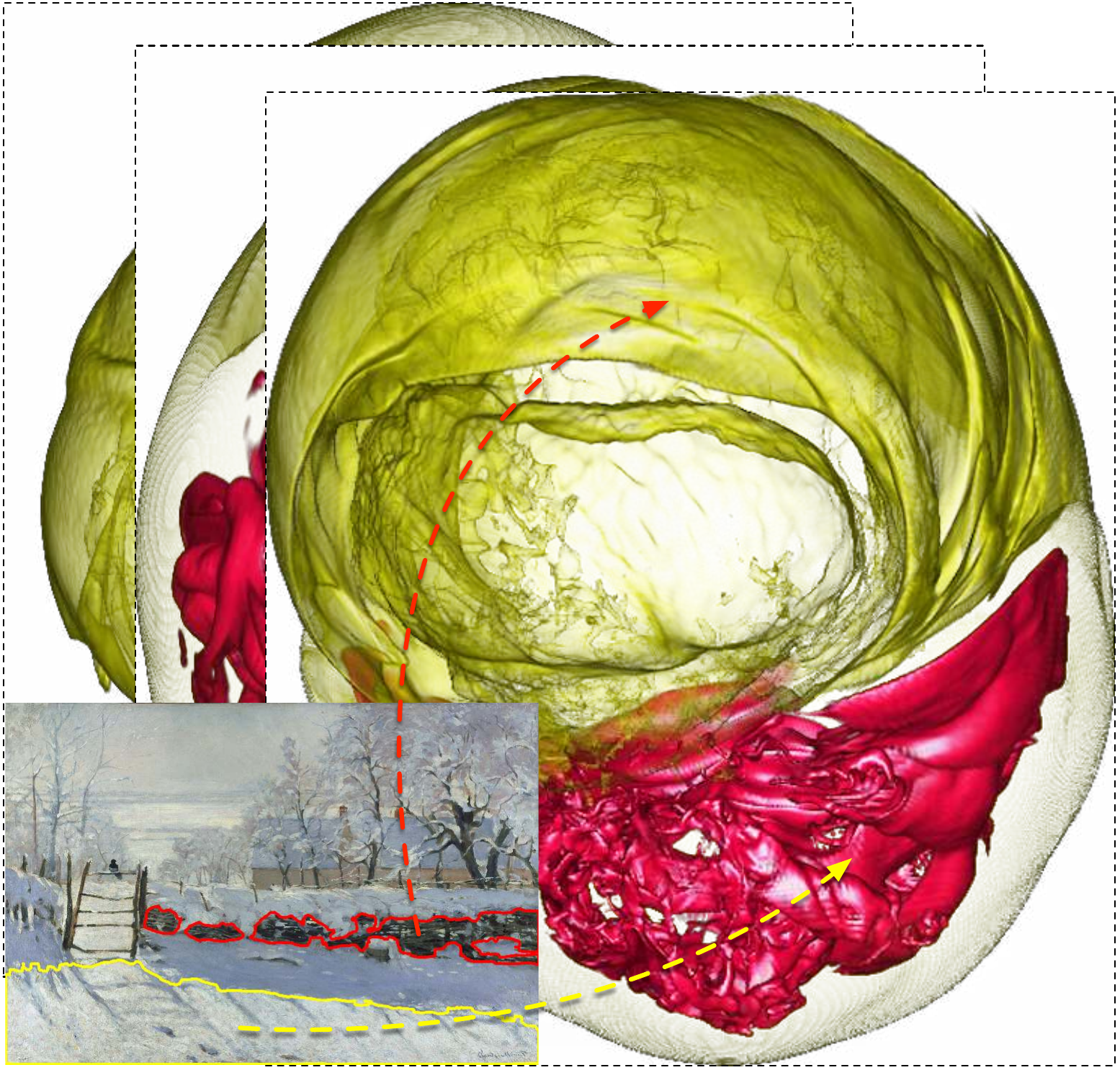} &
		\includegraphics[height=1.2in]{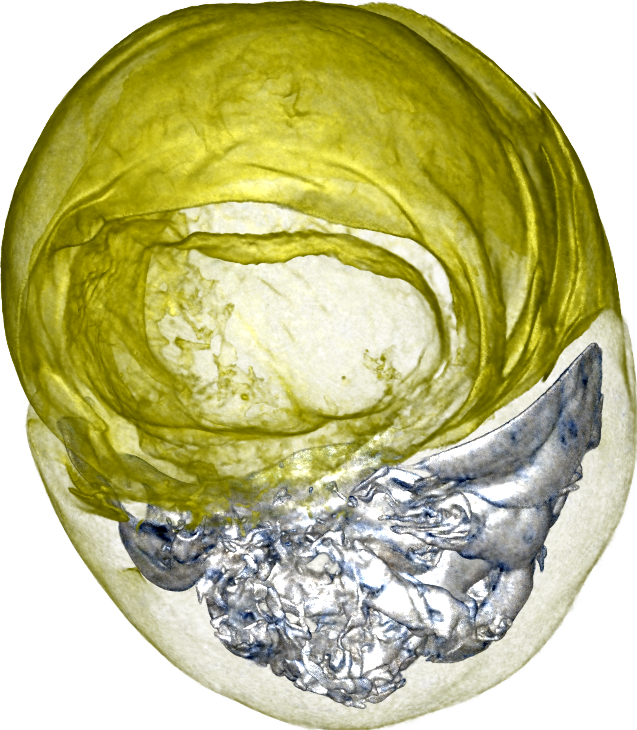} &
		\includegraphics[height=1.2in]{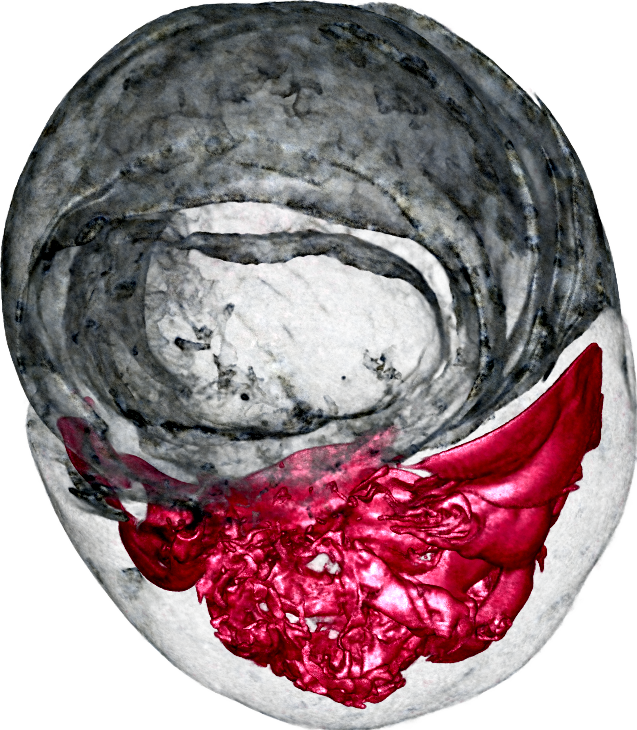} \\
		\mbox{\footnotesize (a) DVR scene and styles}
		& \mbox{\footnotesize (b) partial stylization 1}
		& \mbox{\footnotesize (c) partial stylization 2}
\end{array}$
\end{center}
\vspace{-.25in}
\caption{StyleRF-VolVis partial stylization results on the supernova dataset.} 
\label{fig:PartialStylization}
\end{figure}

{\bf Partial stylization.}
In some scenarios, users may wish to stylize only certain regions of the scene while keeping others unchanged.
Unlike previous NeRF-based stylization methods, StyleRF-VolVis achieves such partial stylization by combining the color representations of PCN and UCN. 
After completing the NPSE of each region, PCN can still utilize the original palette colors to represent the DVR scene. 
Users can thus specify which regions should maintain the NPSE texture and which should keep the original appearance. 
During rendering, regions that preserve the original scene's appearance utilize colors represented by PCN, while those stylized regions are processed through UCN.
Figure~\ref{fig:PartialStylization} shows an example of partial stylization, providing extra freedom for users to achieve desirable results.

\begin{figure}[htb]
\begin{center}
\includegraphics[width=0.95\linewidth]{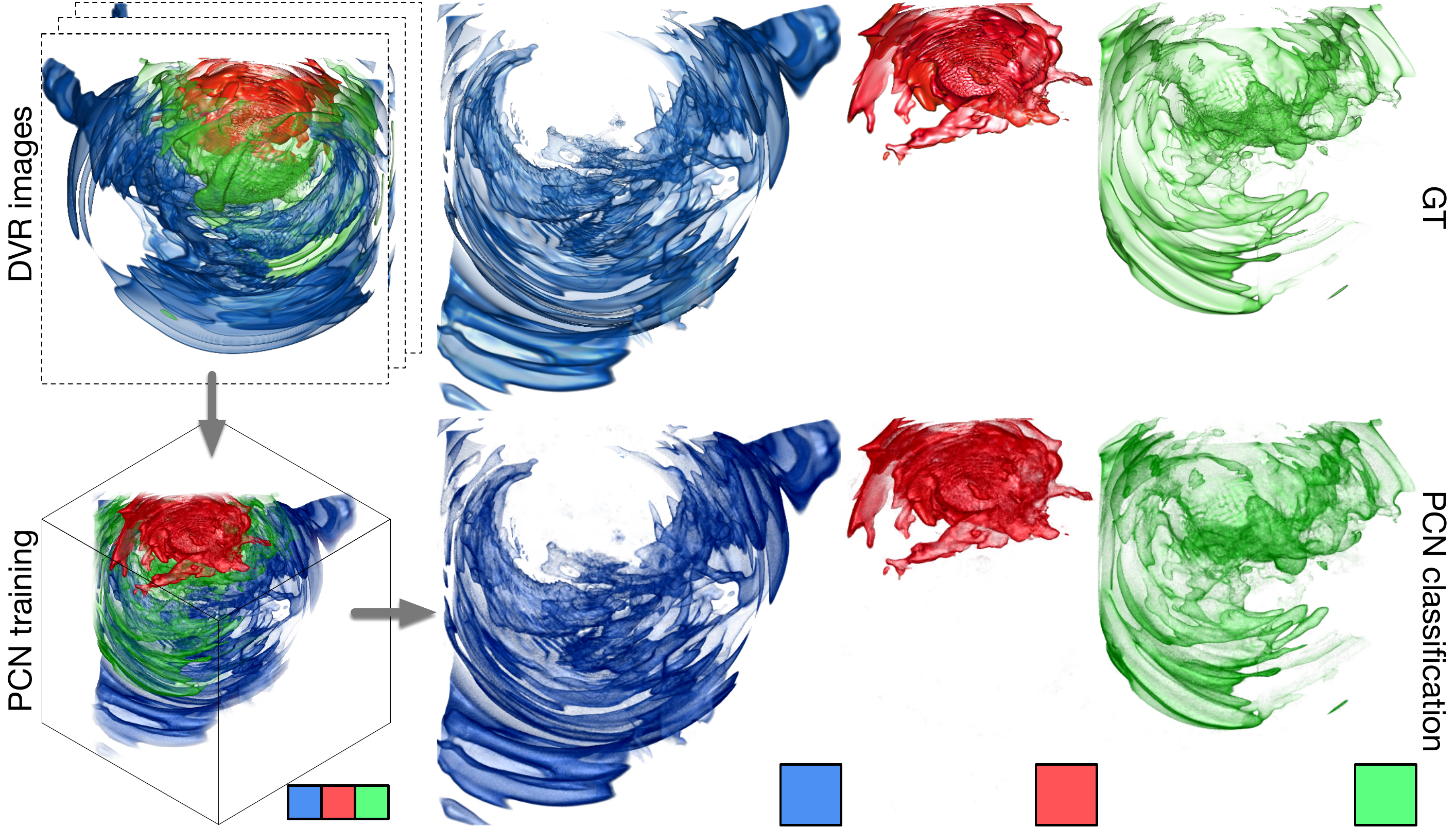}
\end{center}
\vspace{-.25in}
\caption{Comparision of StyleRF-VolVis PCN classification rendering results with GT for each region of the earthquake dataset.} 
\label{fig:PCN-Classification}
\end{figure}

\vspace{-0.05in}
\subsection{PCN Classification and PSE}

As discussed in Section~\ref{sec:NPSE}, the PCN classification outcomes are crucial for style editing.
Visual contents in a DVR scene are often nested with each other. 
However, as demonstrated in Figure~\ref{fig:PCN-Classification}, even for a complex scene like the earthquake, PCN can still produce region classification closely resembling GT, even though only DVR images are used for training.
Although the colors and lighting of PCN classification results are not perfect, the boundary of each region is clear, and even fine details maintain consistency with GT. 
This ensures precise PSE and NPSE of each region in subsequent stages.

\begin{figure}[htb]
\begin{center}
\includegraphics[width=1.00\linewidth]{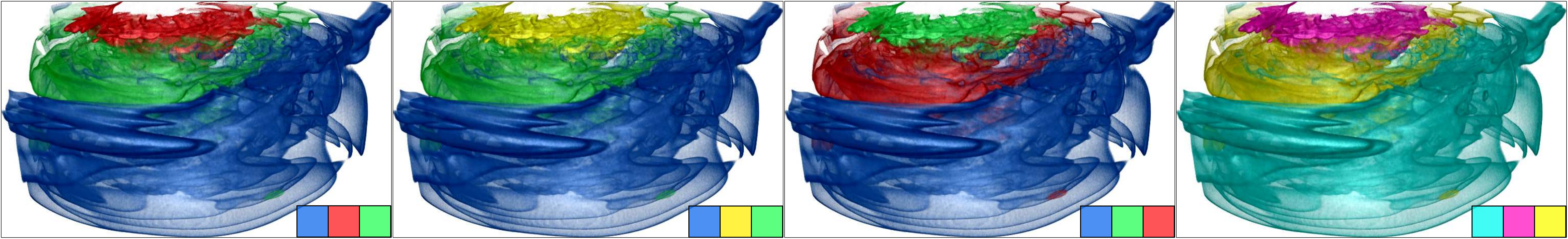}
\includegraphics[width=1.00\linewidth]{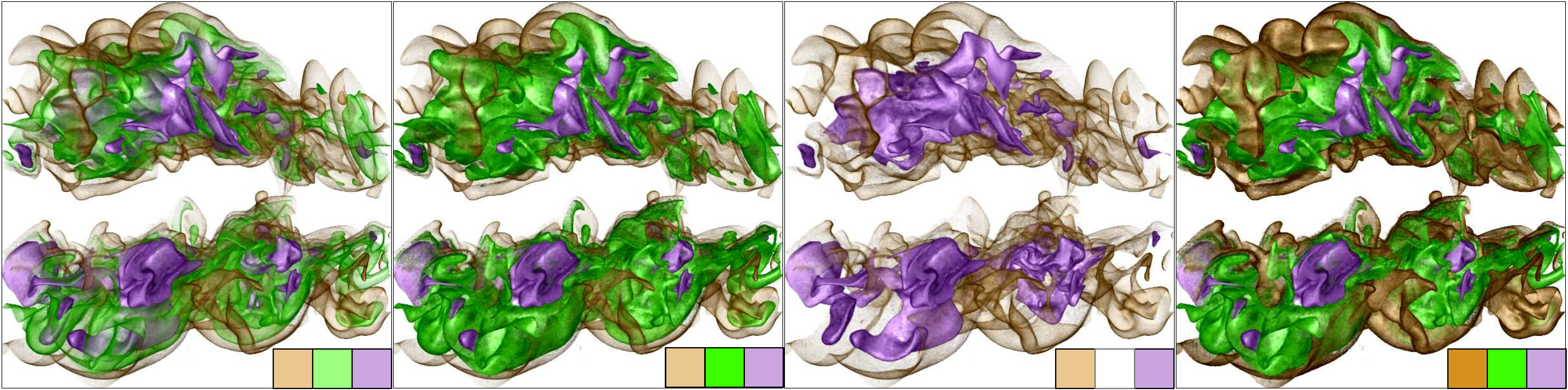}
\includegraphics[width=1.00\linewidth]{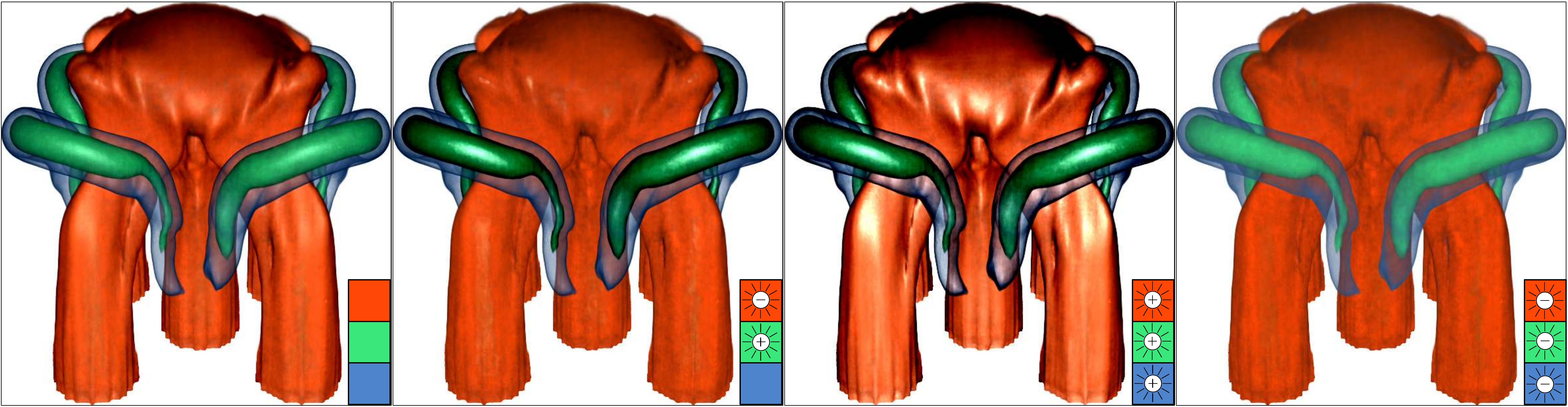}
\end{center}
\vspace{-.25in}
\caption{StyleRF-VolVis PSE of color, opacity, and lighting on earthquake, combustion, and five jets datasets, respectively.} 
\label{fig:PSE-editing}
\end{figure}

In Figure~\ref{fig:PSE-editing}, we showcase the PSE outcomes for various DVR scenes. With the assistance of PCN classification, we can perform a distinct PSE in each region. Such flexibility further assists users in achieving desired stylization results.

\vspace{-0.05in}
\subsection{Ablation Study}

\begin{figure}[htb]
\begin{center}
$\begin{array}{c@{\hspace{0.05in}}c@{\hspace{0.05in}}c}
		\includegraphics[height=1.6in]{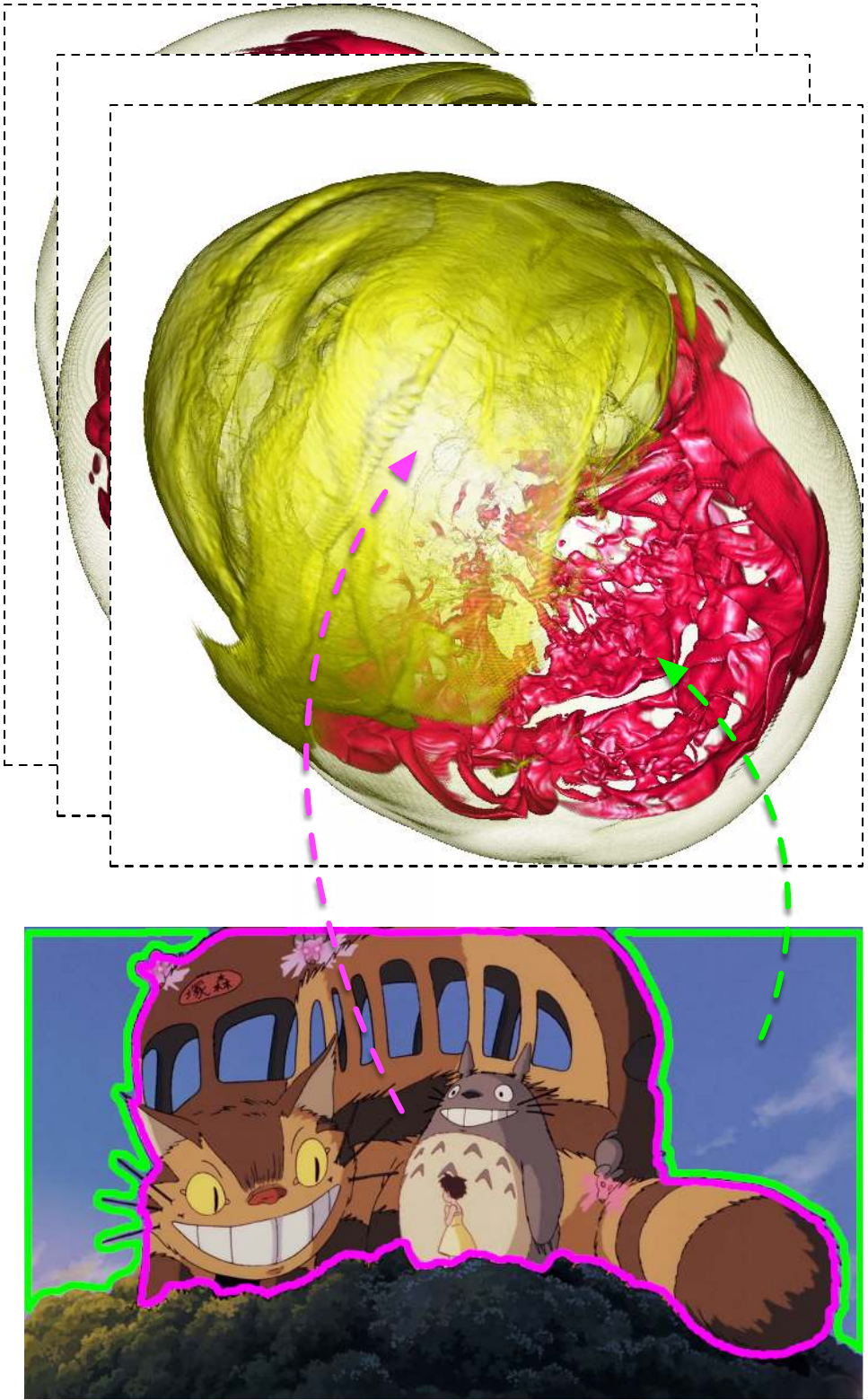} &
		\includegraphics[height=1.6in]{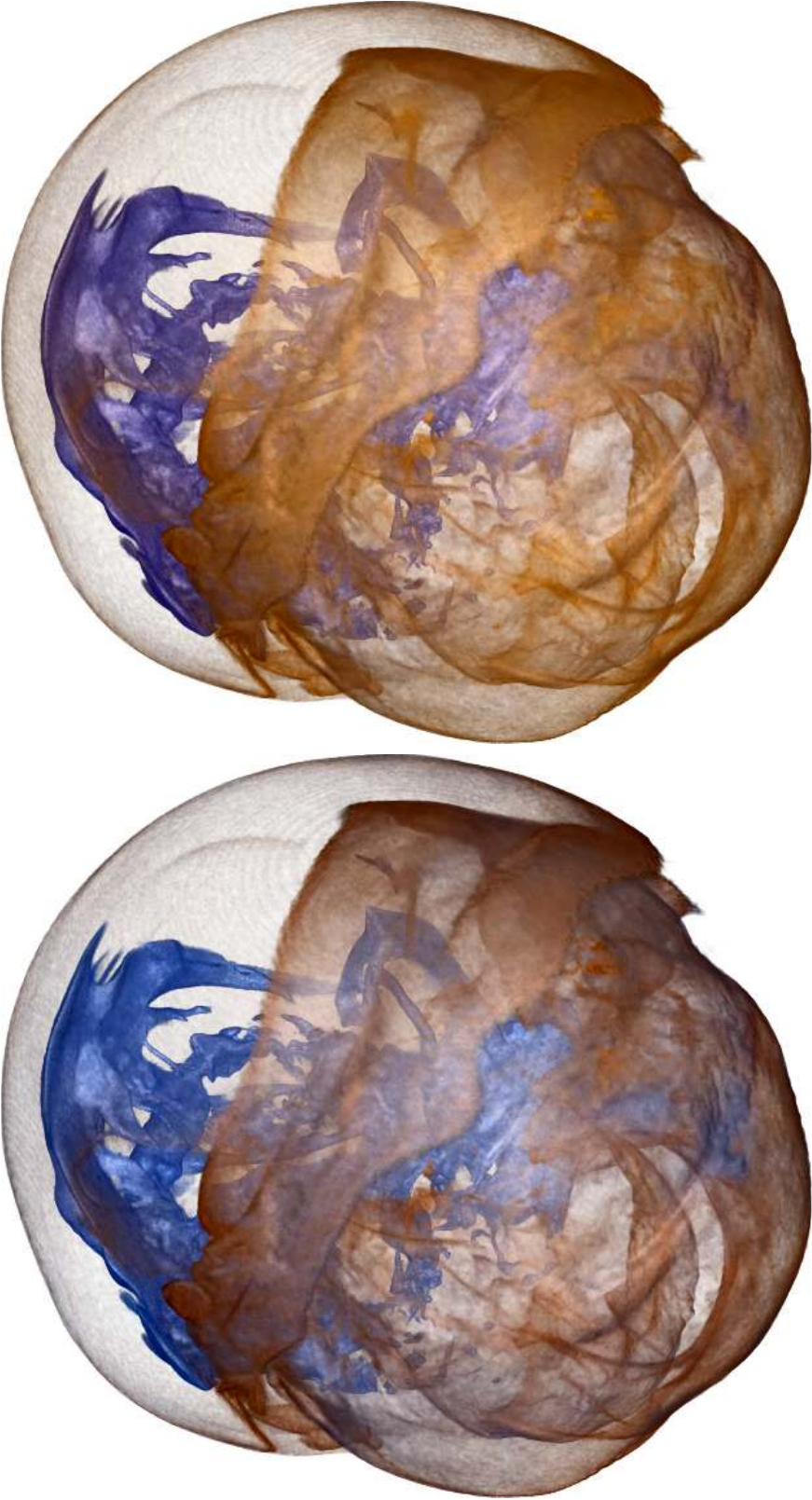} &
		\includegraphics[height=1.6in]{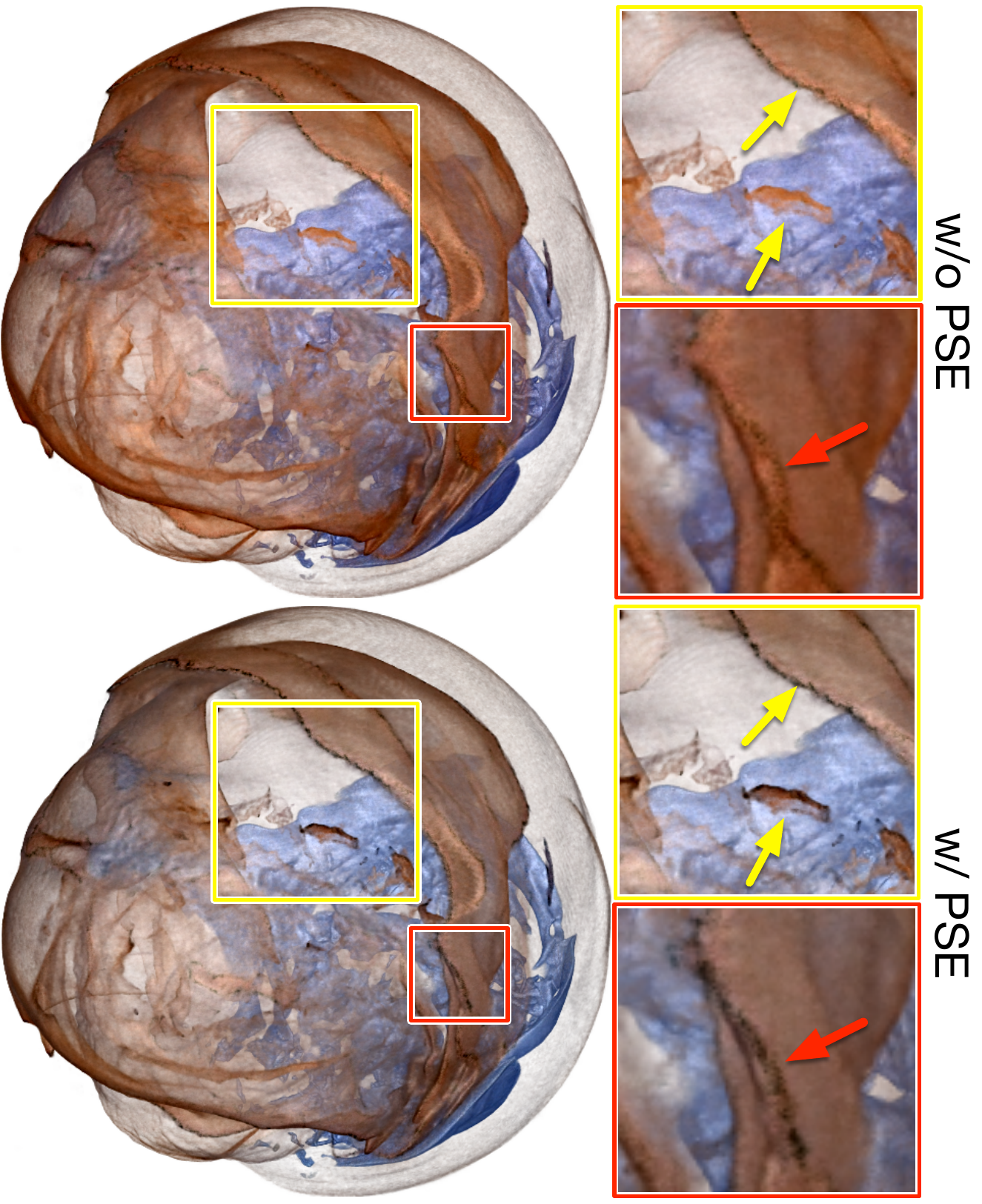} \\
		\mbox{\footnotesize (a) DVR scene and styles}
		& \mbox{\footnotesize (b) 42 iterations}
		& \mbox{\footnotesize (c) 168 iterations}
\end{array}$
\end{center}
\vspace{-.25in}
\caption{Comparison of the style transfer process of StyleRF-VolVis without and with PSE for color transfer on the supernova dataset.} 
\label{fig:PSE4NPSE}
\end{figure}

{\bf Benefit of PSE to NPSE.}
At the beginning of UCN optimization, we perform the color transfer 
in PCN (teacher network) to ensure that the color representation of UCN (student network) matches the selected styles before NPSE. 
This PSE step is essential for speeding up the convergence and improving the stylization quality of UCN during NPSE optimization. 
Figure~\ref{fig:PSE4NPSE} compares the style transfer process without and with color transfer.
Comparing the stylization results of 42 and 168 iterations, we can see that NPSE without color transfer suffers from inaccurate style color matching at an early stage and shows unclear stylization (i.e., black borders, see arrows) at 168 iterations. In contrast, NPSE with color transfer matches the overall style color well at the early stage and reveals well-stylized texture at 168 iterations. 

\begin{figure}[htb]
\begin{center}
$\begin{array}{c@{\hspace{0.05in}}c@{\hspace{0.05in}}c}
		\includegraphics[width=0.37\linewidth]{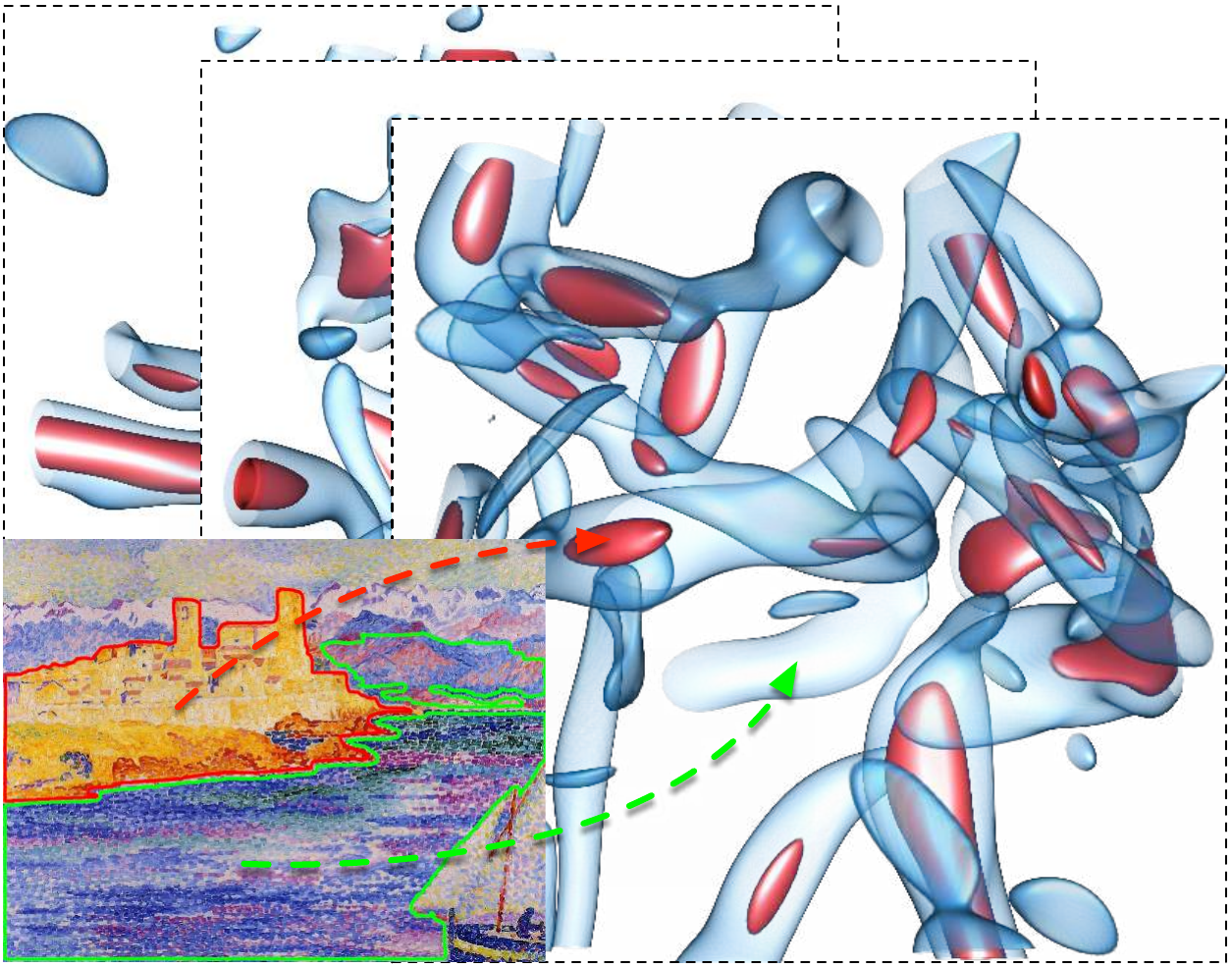} &
		\includegraphics[width=0.28\linewidth]{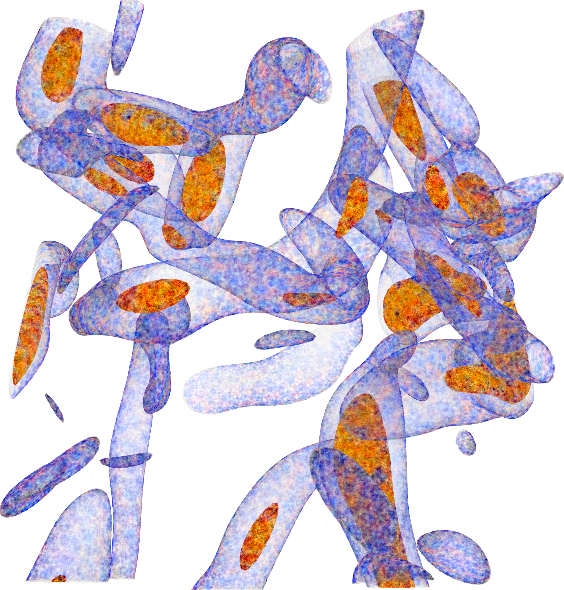} &
		\includegraphics[width=0.28\linewidth]{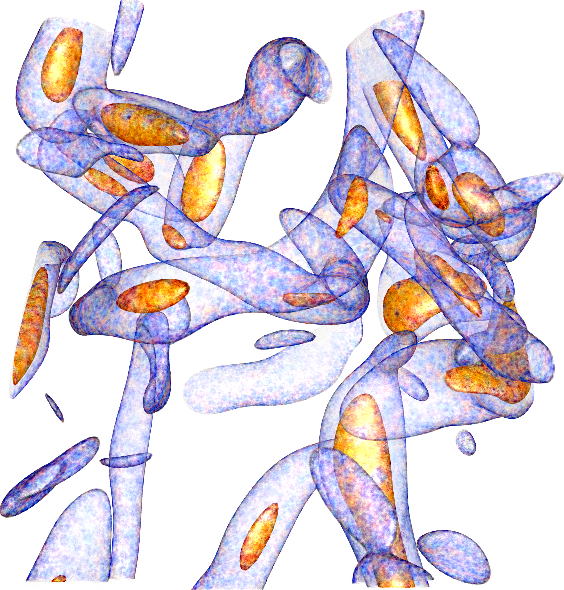} \\
		\includegraphics[width=0.37\linewidth]{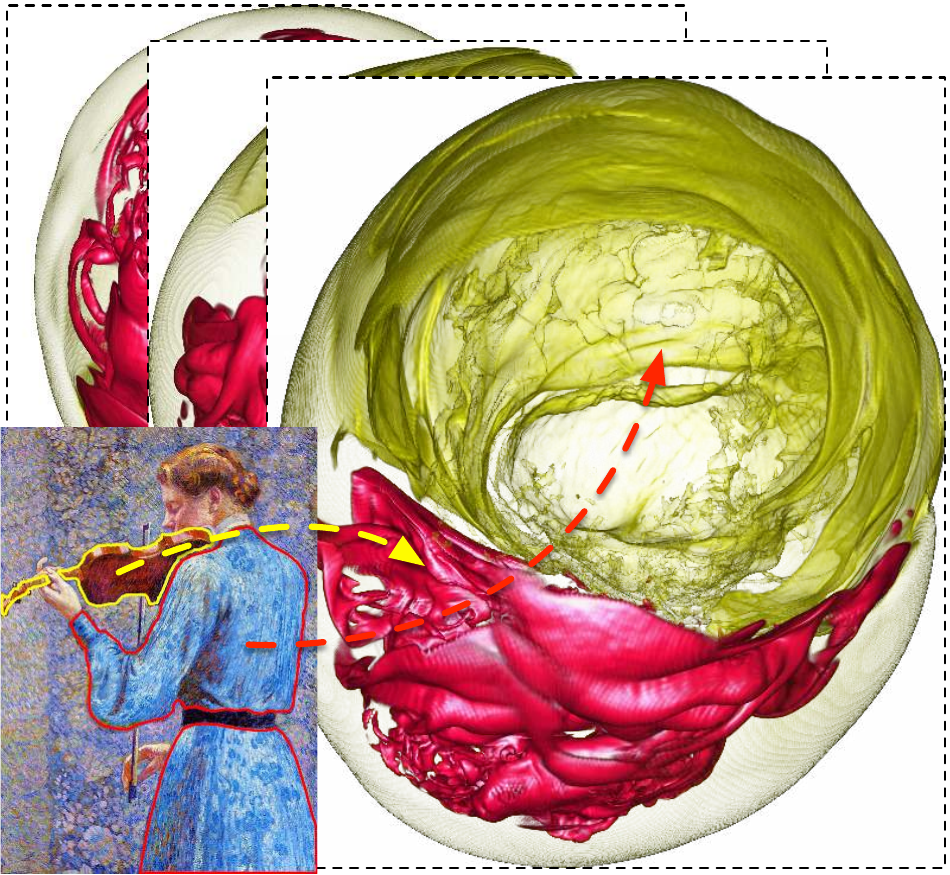} &
		\includegraphics[width=0.28\linewidth]{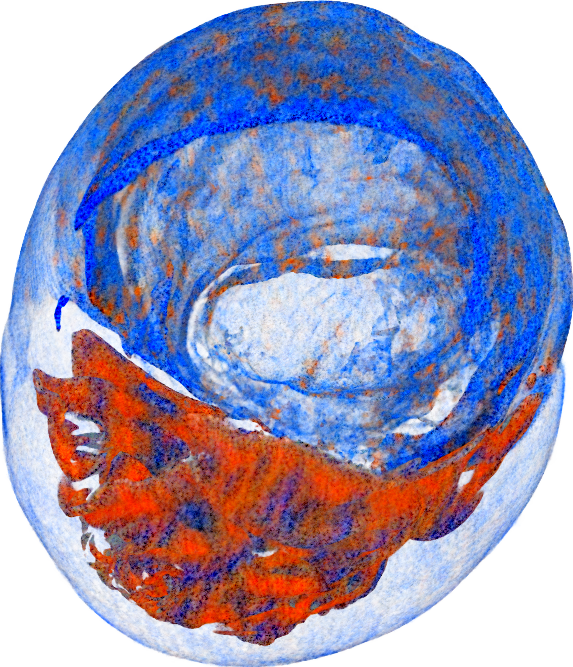} &
		\includegraphics[width=0.28\linewidth]{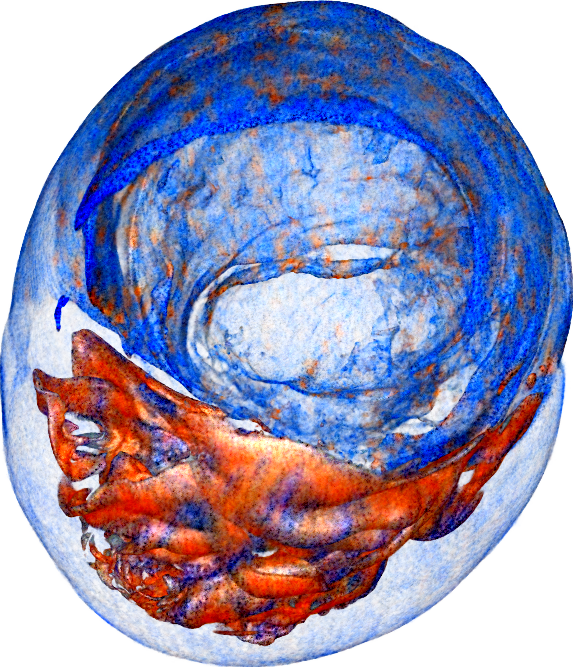} \\
		\mbox{\footnotesize (a) DVR scene and styles}
		& \mbox{\footnotesize (b) w/o lighting}
		& \mbox{\footnotesize (c) w/ lighting}
\end{array}$
\end{center}
\vspace{-.25in}
\caption{Comparison of the stylization results of StyleRF-VolVis without and with lighting on vortex and supernova datasets.} 
\label{fig:lighting4NPSE}
\end{figure}

{\bf View-dependent lighting in NPSE.} 
Recent NeRF-based stylization methods~\cite{Chiang-WACV22,Huang-CVPR22,Zhang-ARF,Nguyen-Phuoc-SNeRF,Liu-CVPR23} discard view directions in the input to ensure cross-view stylization consistency. 
However, they ignore that view-dependent lighting of the original scene is essential for maintaining consistency between the original content and the stylized scene.
In contrast, StyleRF-VolVis utilizes the additional lighting MLP optimized in the PCN training stage to preserve the DVR lighting during NPSE. 
In Figure~\ref{fig:lighting4NPSE}, we show examples of vortex and supernova datasets to compare the effect of DVR lighting on NPSE outcomes. 
The results show that the stylized scene with lighting is more consistent with the original DVR scene than that without. 

\vspace{-0.05in}
\subsection{Limitations}

Although StyleRF-VolVis achieves flexible, high-quality stylization of DVR scenes, it has the following limitations. 
\pin{First, given an initial TF, NeRF cannot represent the value ranges where opacity equals zero. Consequently, StyleRF-VolVis cannot perform PSE or NPSE for regions with zero opacity in the initial TF.}  
Second, StyleRF-VolVis supports interactive PSE but not interactive NPSE.
Third, even though users can adjust the lighting {\em intensity} for the DVR scene, they cannot modify the lighting {\em direction} because no normal information is available.
Finally, 
if the colors among different regions within the input DVR scene are similar, color refinement may not separate these regions correctly. In such cases, users may need to manually adjust the number of palette colors and their RGB values to obtain the desired results.

\vspace{-0.05in}
\section{Conclusions and Future Work}

We have presented StyleRF-VolVis, the first work in VolVis that targets style transfer in the NeRF space. 
The crux of our approach lies in the accurate extraction of content and appearance information separately from the given DVR scene and the bridging between photorealistic and non-photorealistic style editing via knowledge distillation. 
With these innovations, StyleRF-VolVis achieves high-quality, consistent, and flexible 3D style transfer outcomes with novel view synthesis. 
The efficacy of StyleRF-VolVis is demonstrated with various combinations of DVR scenes and reference images. 
Moreover, we compare StyleRF-VolVis against other image-based (AdaIN), video-based (ReReVST), and NeRF-based (ARF and SNeRF) solutions via objective and subjective evaluation to showcase its superior quality performance. 

In the future, we will extend StyleRF-VolVis to handle dynamic DVR scenes produced from time-varying datasets. 
The challenge is maintaining a consistent appearance over timesteps to achieve temporally coherent stylization. 
We will also explore StyleRF-VolVis for multivariate or ensemble VolVis, where different variables or ensembles could be mapped to visually distinct styles beyond colors for better differentiation. 
It remains to be seen what the appropriate number of styles and their mixing should be to leverage the human's visual capacity best while maintaining observation clarity.

To ease the difficulty for non-professionals using StyleRF-VolVis, we will explore integrating natural language interaction into the current graphical user interface and broaden the selection of reference images from the WikiArt collection to images created by generative AI. 
The success of StyleRF-VolVis will unfold exciting opportunities for VolVis beyond the originated scientific domain. We envision that such a solution will enable citizen science by fusing diverse disciplines, such as science and art, for the general public's exploration, understanding, and appreciation, which we would like to pursue.

\vspace{-0.05in}
\acknowledgments{This research was supported in part by the U.S.\ National Science Foundation through grants IIS-1955395, IIS-2101696, OAC-2104158, and IIS-2401144, and the U.S.\ Department of Energy through grant DE-SC0023145. The authors would like to thank the anonymous reviewers for their insightful comments.}

\vspace{-0.05in}
\bibliographystyle{abbrv-doi-hyperref}

\bibliography{template}

\newpage

\section*{Appendix}
\setcounter{section}{0}
\setcounter{figure}{0}
\setcounter{table}{0}

\section{Implementation Details}

{\bf Palette refinement algorithm.}
When initializing palette colors $\textbf{P}$ before PCN training, we refine $\textbf{P}_{\convex}$ extracted from the RGB convex hull method
to eliminate similar colors within the palette. 
Specifically, for any two colors in $\textbf{P}_{\convex}$, if they are similar (i.e., their L1 distance of {\em hue} value in the HSB space is below a threshold $T_h$), we remove one of them. 
Furthermore, there is a small chance that $\textbf{P}_{\convex}$ may contain gray colors corresponding to the light color in the DVR scene. 
Such gray colors are unnecessary in $\textbf{P}$ as the lighting MLP has represented lighting components. 
Therefore, we remove any color in $\textbf{P}_{\convex}$ if its {\em brightness} value in the HSB space is less than a threshold $T_b$. 
For normalized color component values in the HSB space, we empirically set $T_h=0.1$ and $T_b=0.2$ to obtain refined $\textbf{P}$.

{\bf Luminance background.}
Before NPSE, we calculate each style's luminance value $L$ to compute the NNFM loss using VGG-16.
Following the calculation steps given in~\cite{stone2003field}, for each average RGB color of the selected style, we normalize each component, such as $R$, and convert it into the linear-scale counterpart $R_{\lin}$
\begin{equation}
	R_{\lin} =  R^{2.2}.
\end{equation}
$L$ of the selected style can be computed as
\begin{equation}
	L = 0.2126\times R_{\lin} + 0.7152\times G_{\lin} + 0.0722\times B_{\lin},
\end{equation}
where the RGB component coefficients reflect the average spectral sensitivity of lighting perceived by humans. 
We use $L$ as the corresponding background color for each style rendering. 
This way, the stylization of StyleRF-VolVis can better match the overall brightness of the selected style.

\begin{figure}[htb]
\begin{center}
$\begin{array}{c@{\hspace{0.05in}}c@{\hspace{0.05in}}c}
		\includegraphics[width=0.3\linewidth]{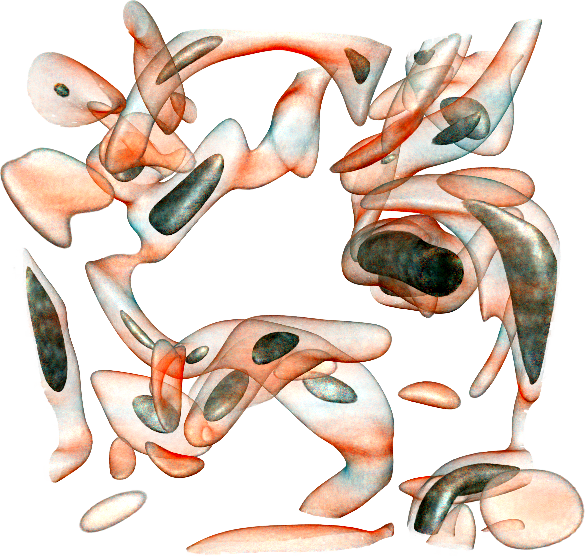} &
		\includegraphics[width=0.3\linewidth]{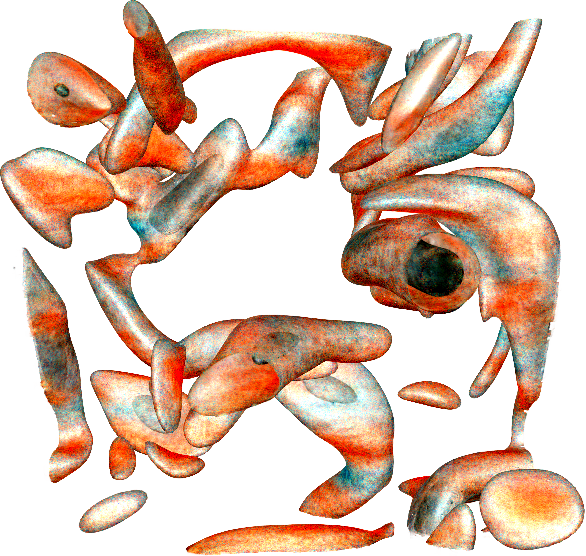} &
		\includegraphics[width=0.3\linewidth]{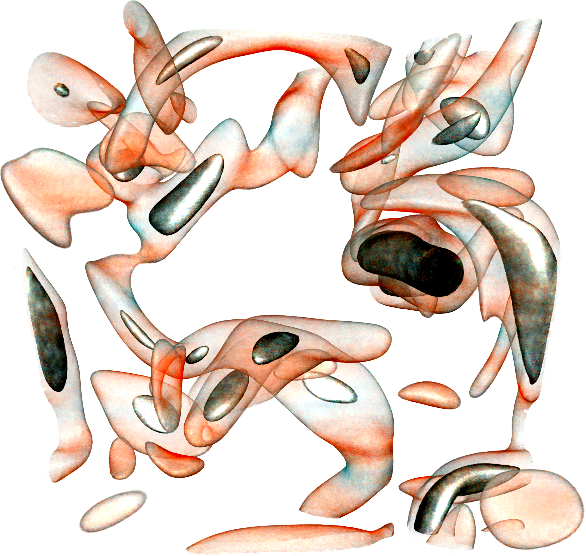} \\
		\includegraphics[width=0.3\linewidth]{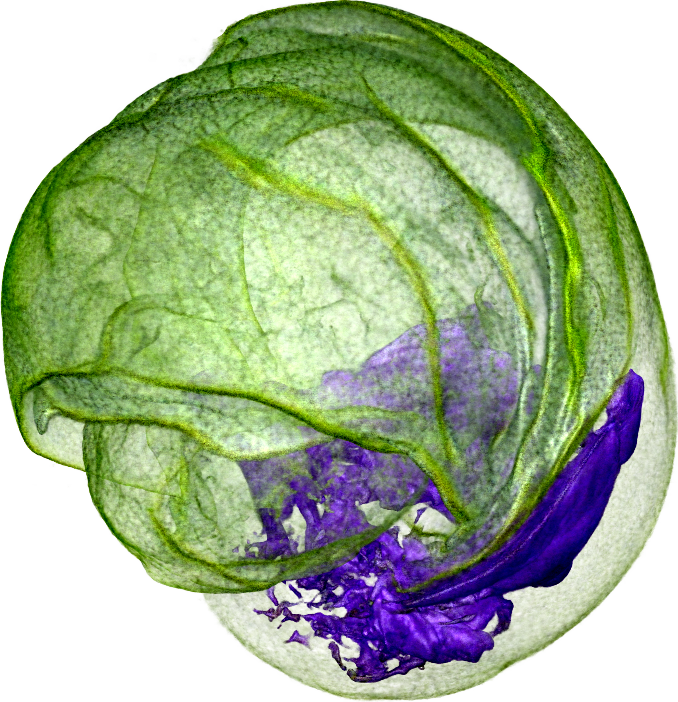} &
		\includegraphics[width=0.3\linewidth]{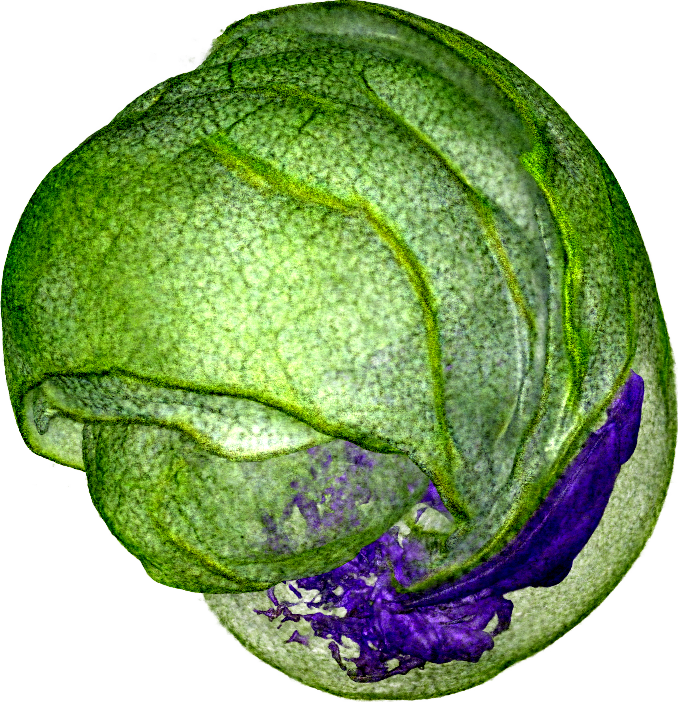} &
		\includegraphics[width=0.3\linewidth]{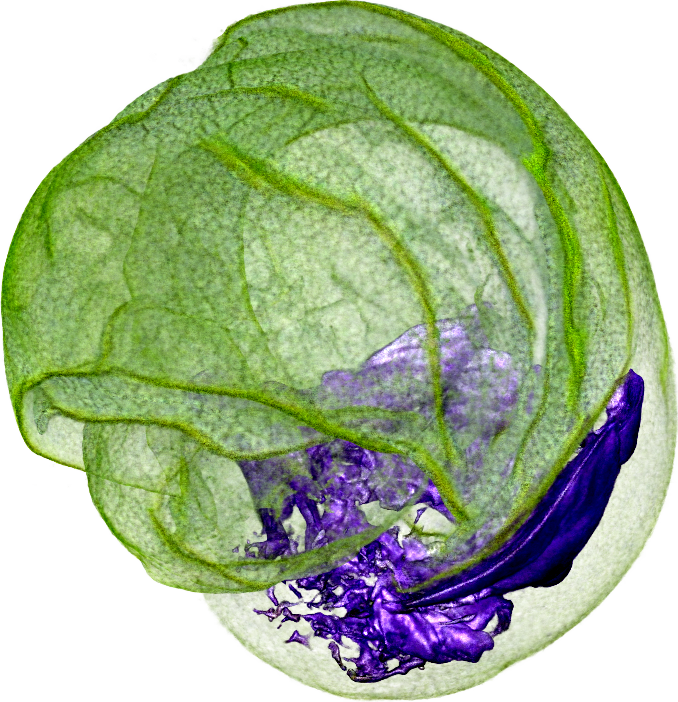} \\
		\includegraphics[width=0.3\linewidth]{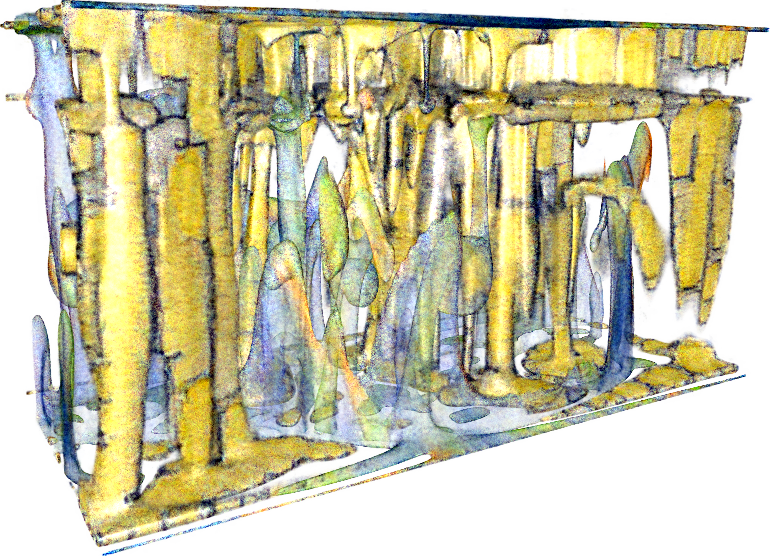} &
		\includegraphics[width=0.3\linewidth]{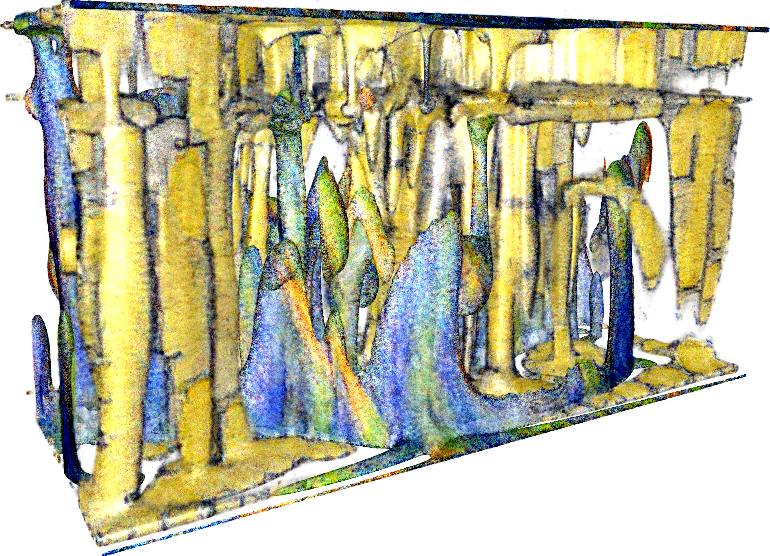} &
		\includegraphics[width=0.3\linewidth]{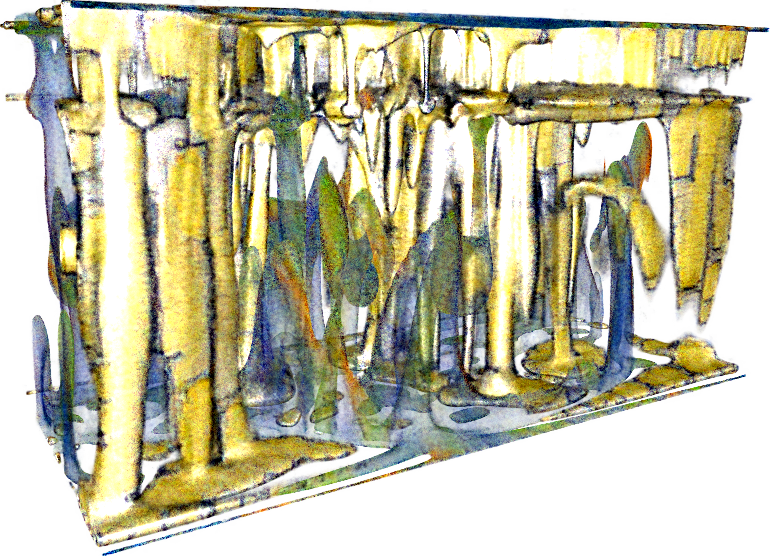} \\
		\mbox{\footnotesize (a) NPSE scene}
		& \mbox{\footnotesize (b) opacity PSE}
		& \mbox{\footnotesize (c) lighting PSE}
\end{array}$
\end{center}
\vspace{-.25in}
\caption{Applying various opacity or lighting PSEs to the stylized NPSE scenes.} 
\label{fig:PSEafterNPSE}
\end{figure}

\section{Additional Results}

{\bf PSE after NPSE.}
After NPSE, users can still apply further PSE to the stylized scene. 
When doing so, one limitation is that if a region of the NPSE scene utilizes the UCN to represent the color term, users cannot apply a color PSE to the region as the PCN does not represent the color.
However, users can modify the opacity and lighting of each region without restriction.
Figure~\ref{fig:PSEafterNPSE} shows examples of applying various opacity or lighting PSEs to the scene after NPSE.

\begin{table}[htb]
\caption{Averaging PSNR (dB), LPIPS, and SSIM values across all PCN-inferred images with the combustion dataset. The best ones are shown in bold.}
\vspace{-0.1in}
\centering
{\scriptsize
\begin{tabular}{c|ccc}
$\lambda_\delta$ & PSNR$\uparrow$  &LPIPS$\downarrow$ &SSIM$\uparrow$ \\ \hline
0.0      		 &20.47    &0.079   &0.954\\
0.1 			 &\textbf{23.93}    &\textbf{0.056}  	&\textbf{0.969}\\
1.0 			 &22.99    &0.058   &0.967\\ 
\end{tabular}
}
\label{tab:comp-delta}
\end{table}

\begin{figure}[htb]
\begin{center}
$\begin{array}{c@{\hspace{0.05in}}c}
		\includegraphics[width=0.45\linewidth]{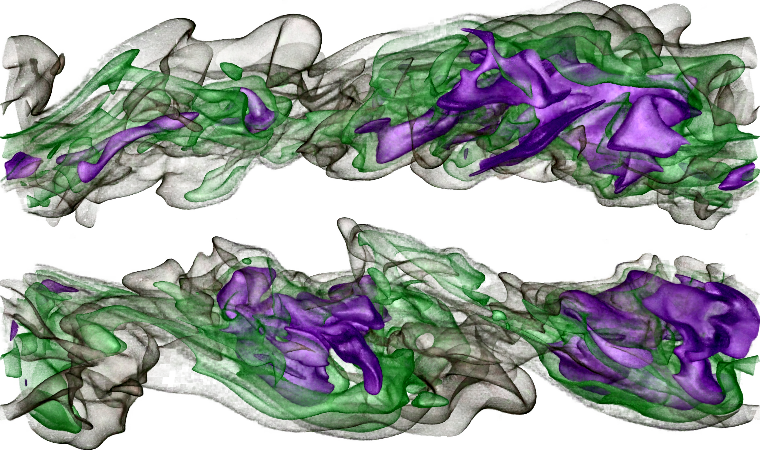}  &
		\includegraphics[width=0.45\linewidth]{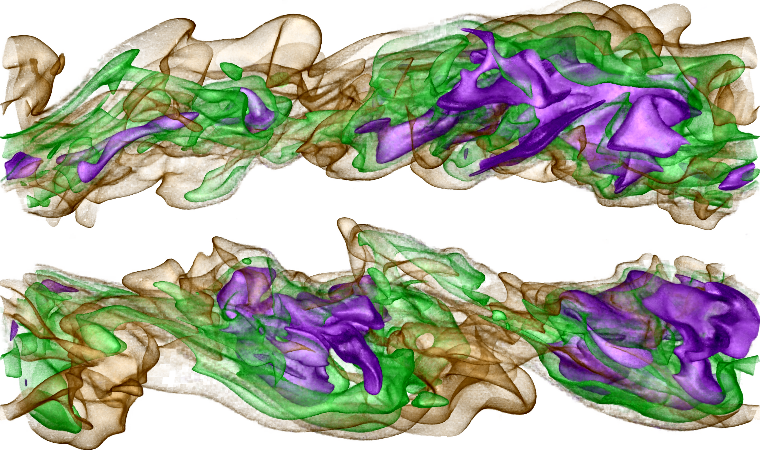}  \\
		\mbox{\footnotesize (a) $\lambda_\delta$ = 0.0}
		& \mbox{\footnotesize (b) $\lambda_\delta$ = 1.0}\\
		\includegraphics[width=0.45\linewidth]{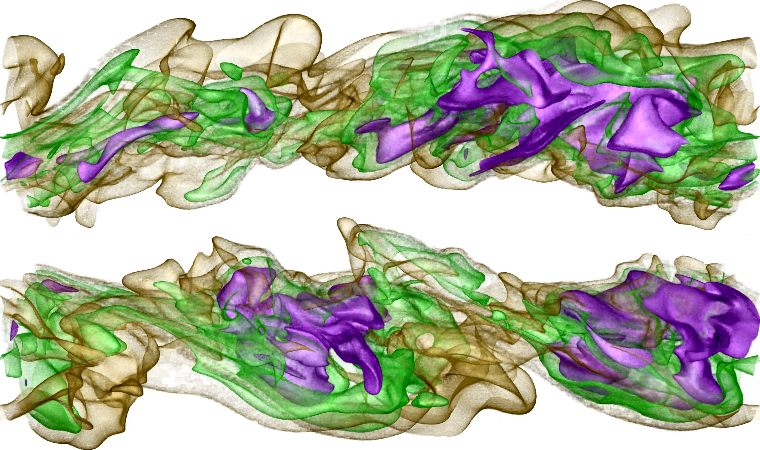} &
		\includegraphics[width=0.45\linewidth]{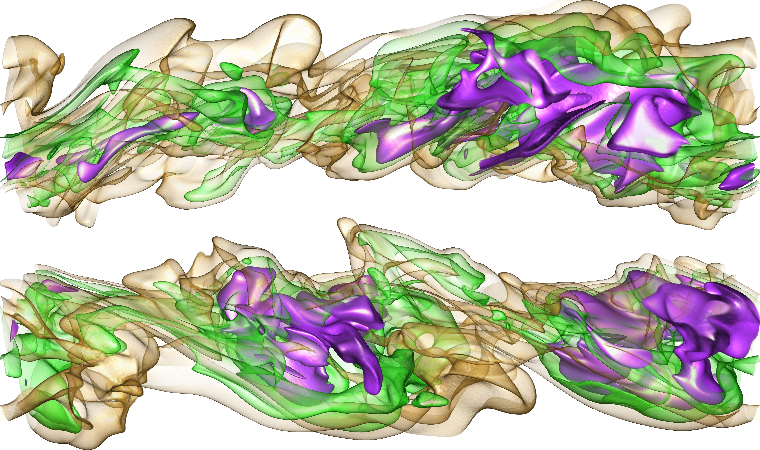}  \\
		\mbox{\footnotesize (c) $\lambda_\delta$ = 0.1}
		& \mbox{\footnotesize (d) GT}
\end{array}$
\end{center}
\vspace{-.25in}
\caption{Comparison of PCN rendering results under different $\lambda_\delta$ on the combustion dataset.} 
\label{fig:comp-lambda}
\end{figure}

\begin{figure}[htb]
\begin{center}
$\begin{array}{c@{\hspace{0.05in}}c}
		\includegraphics[width=0.42\linewidth]{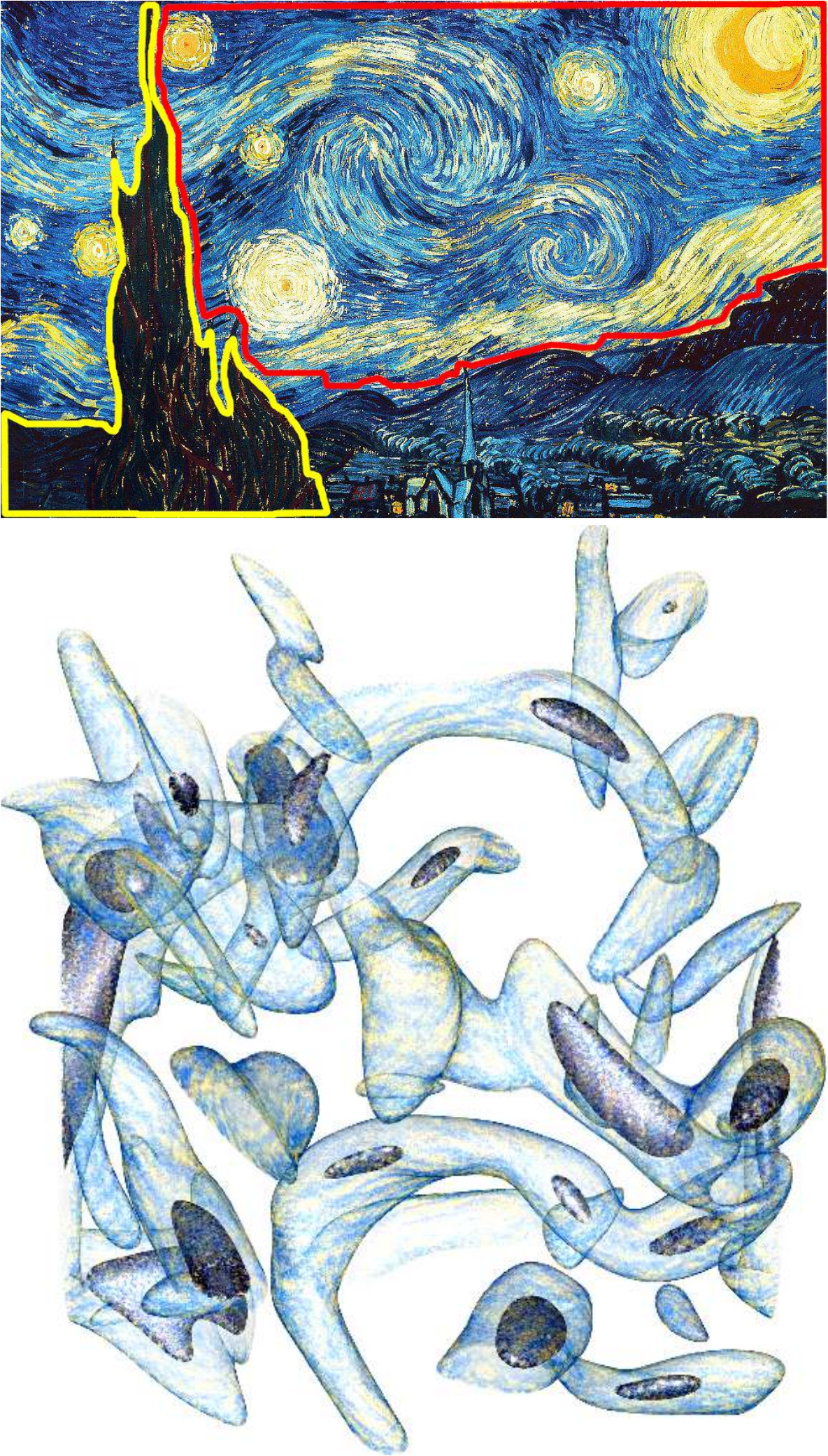}  &
		\includegraphics[width=0.42\linewidth]{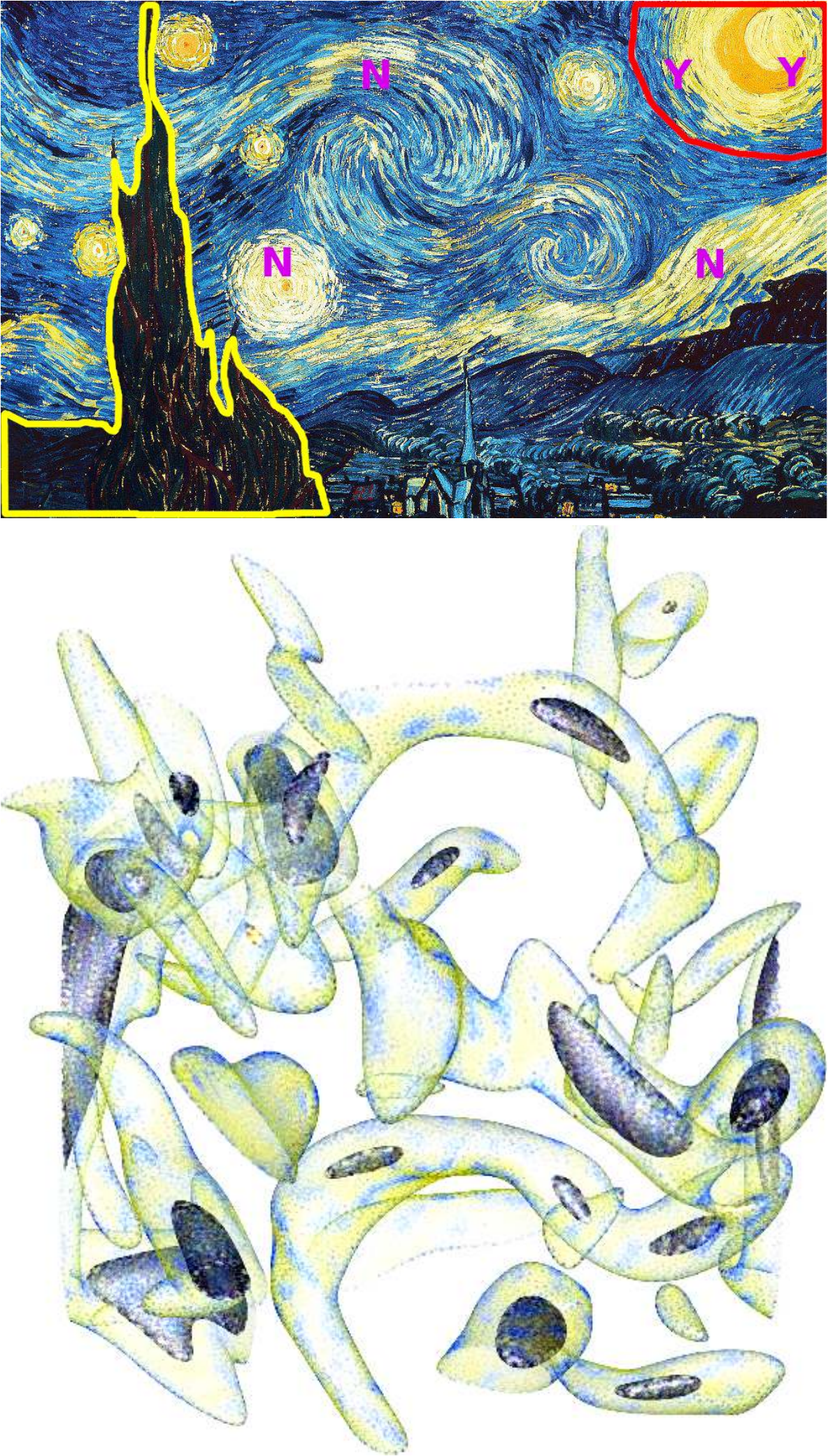}  \\
		\mbox{\footnotesize (a) before refinement}
		& \mbox{\footnotesize (b) after refinement}
\end{array}$
\end{center}
\vspace{-.25in}
\caption{\pin{Iterative style refinement on the vortex dataset. Y/N shows a positive/negative point prompt to include/exclude a certain selection.}} 
\label{fig:style-refinement}
\end{figure}

{\bf Choice of $\lambda_\delta$.}
When optimizing the PCN to avoid palette color shiftings, we include an offset regularization loss and use $\lambda_\delta$ to control the regularization strength. 
We conduct an ablation study to investigate the effect of $\lambda_\delta$ on PCN performance.
After optimizing the PCN with different $\lambda_\delta$, we compare PCN rendering results with GT using PSNR, LPIPS, and SSIM, as shown in Table~\ref{tab:comp-delta}. 
Figure~\ref{fig:comp-lambda} presents the rendering results. 
We can see that the offset regularization loss is essential for PCN training. 
When offset regularization is missing ($\lambda_\delta=0$), the PCN does not predict correctly, as it tries to focus more on leveraging offsets instead of palette colors to represent the DVR scene.
However, the PCN may leverage more on palette colors instead of offsets to represent colors when $\lambda_\delta$ gets larger, resulting in less accurate scene reconstruction. 
Therefore, we choose $\lambda_\delta = 0.1$ for a good control of the regularization strength.

\pin{{\bf Iterative style refinement for NPSE.}
For NPSE, StyleRF-VolVis does not support direct control over the stylization process. 
However, users can indirectly achieve their desired stylization by iterative refining the selection in the reference image.
Figure~\ref{fig:style-refinement} shows such an example.
Based on the style selection in (a), we apply several negative point prompts for SAM in (b) to exclude undesired styles and retrain the UCN to achieve the desired stylization.}

\begin{figure}[htb]
\begin{center}
$\begin{array}{c@{\hspace{0.0in}}c}
		\includegraphics[height=1.5in]{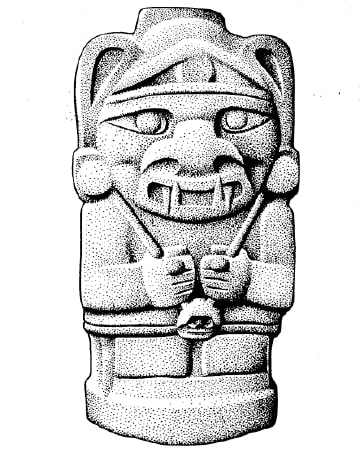}  &
		\includegraphics[height=1.5in]{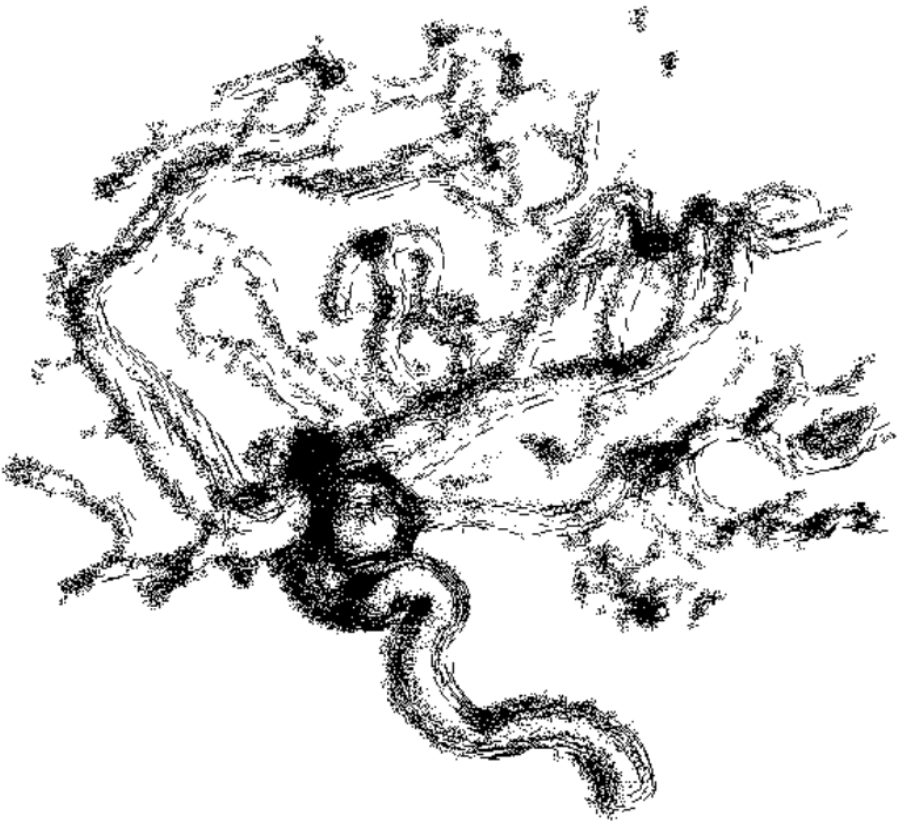}  \\
		\mbox{\footnotesize (a) reference stipple drawing~\cite{Lu-VIS02}}
		& \mbox{\footnotesize (b) stylization from~\cite{Lu-VIS02}}\\
		\includegraphics[width=0.45\linewidth]{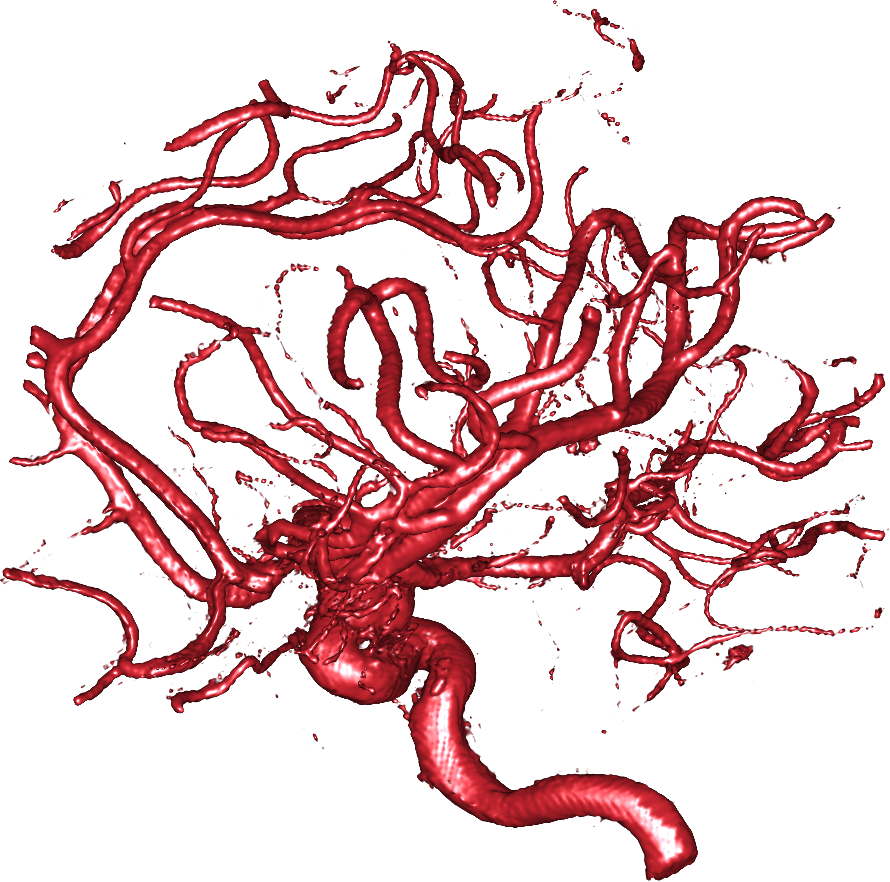} &
		\includegraphics[width=0.45\linewidth]{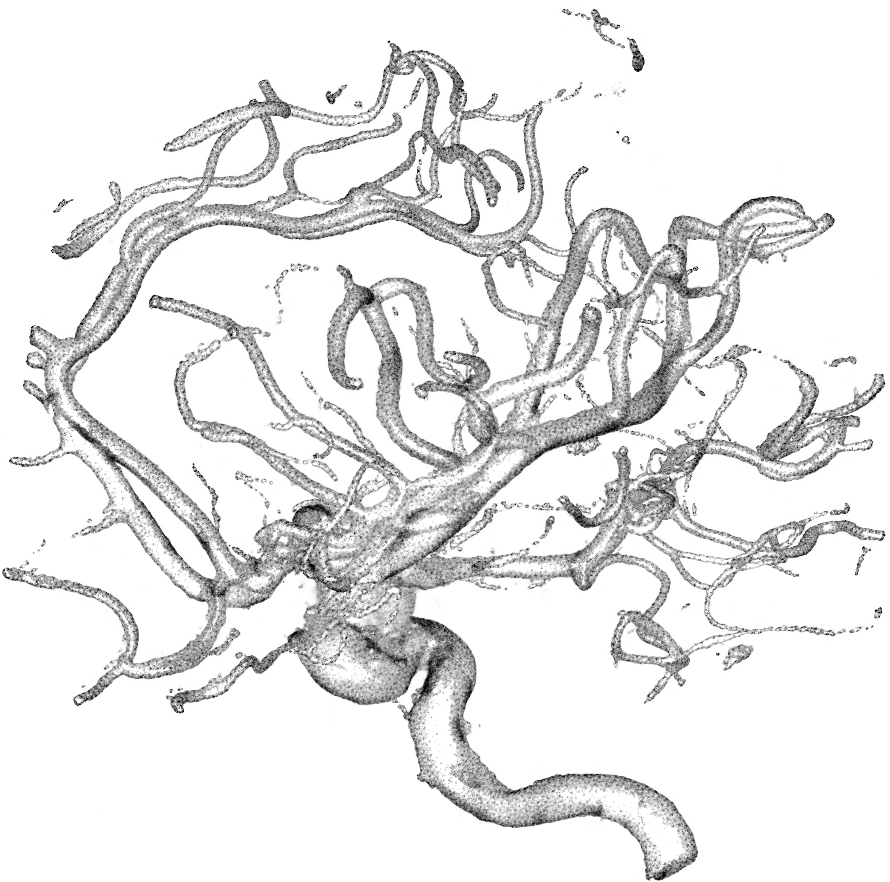}  \\
		\mbox{\footnotesize (c) DVR}
		& \mbox{\footnotesize (d) stylization from StyleRF-VolVis}
\end{array}$
\end{center}
\vspace{-.25in}
\caption{\pin{Comparison of the stylization results generated by a NPR method~\cite{Lu-VIS02} and StyleRF-VolVis on the aneurysm dataset.}} 
\label{fig:Stipple}
\end{figure}

%

\begin{figure}[htb]
\begin{center}
$\begin{array}{c@{\hspace{0.05in}}c}
		\includegraphics[width=0.45\linewidth]{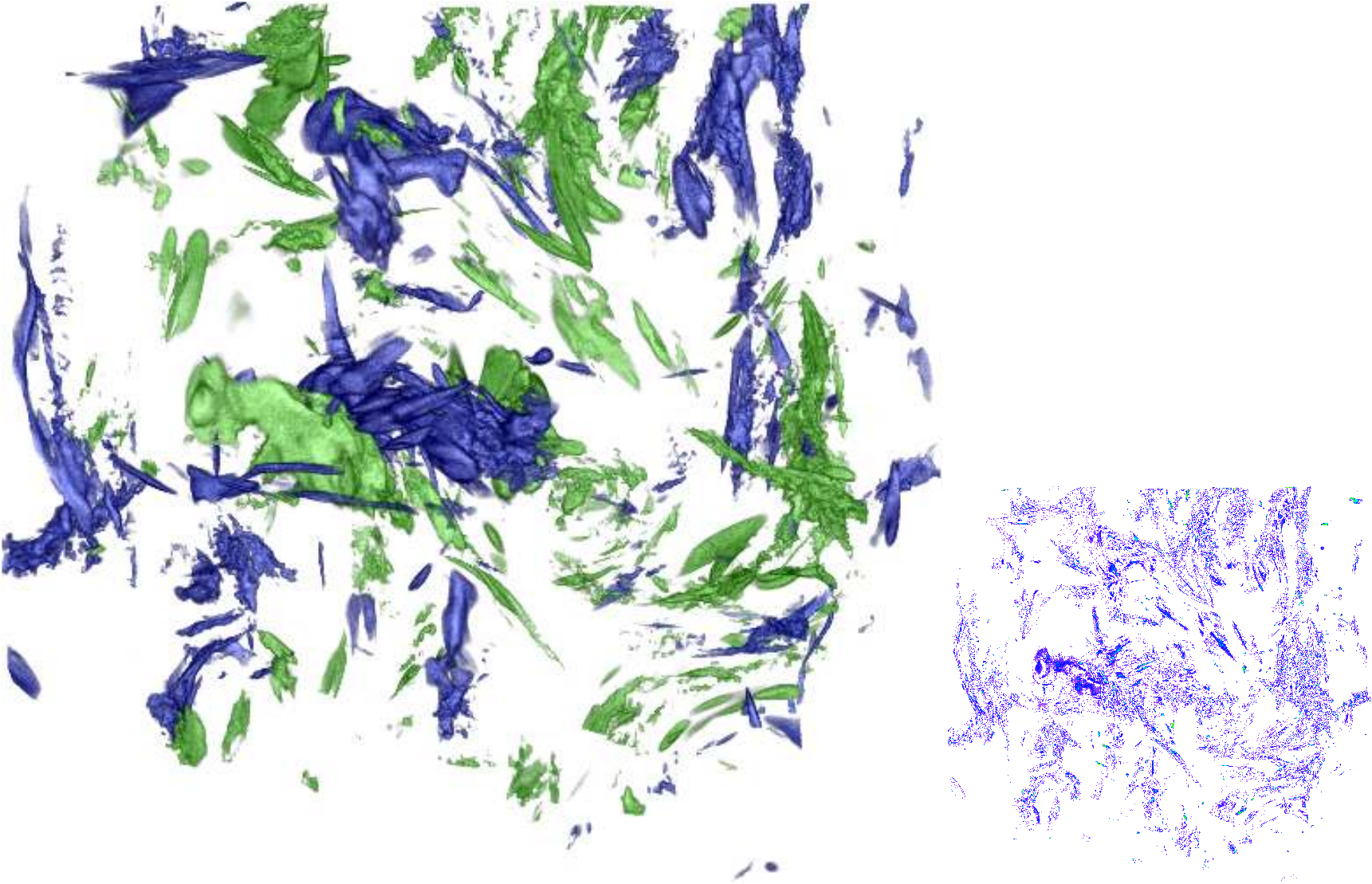}  &
		\includegraphics[width=0.45\linewidth]{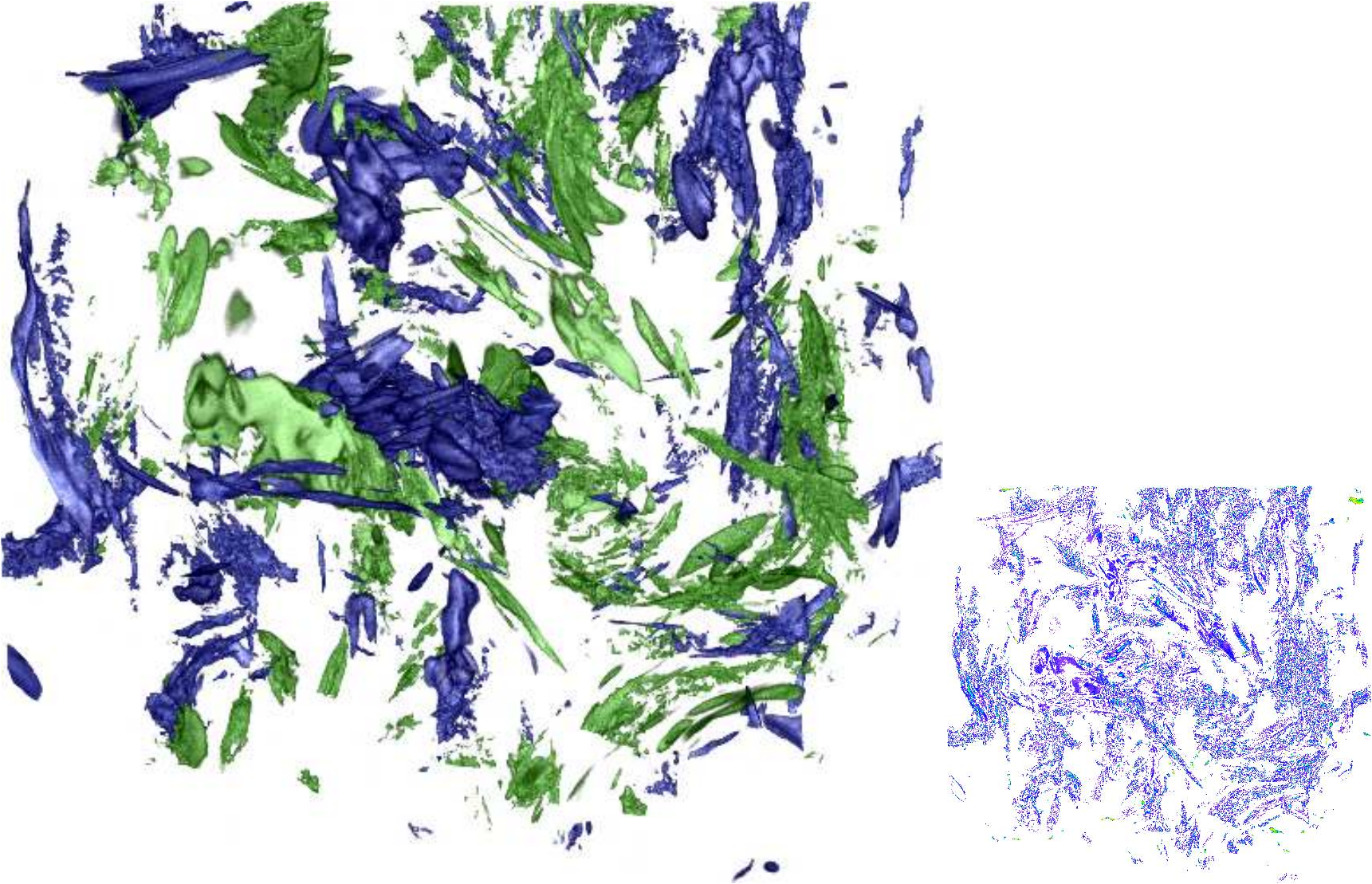}  \\
		\mbox{\footnotesize (a) E1} & \mbox{\footnotesize (b) E3}\\
\end{array}$
$\begin{array}{c}
		\includegraphics[width=0.45\linewidth]{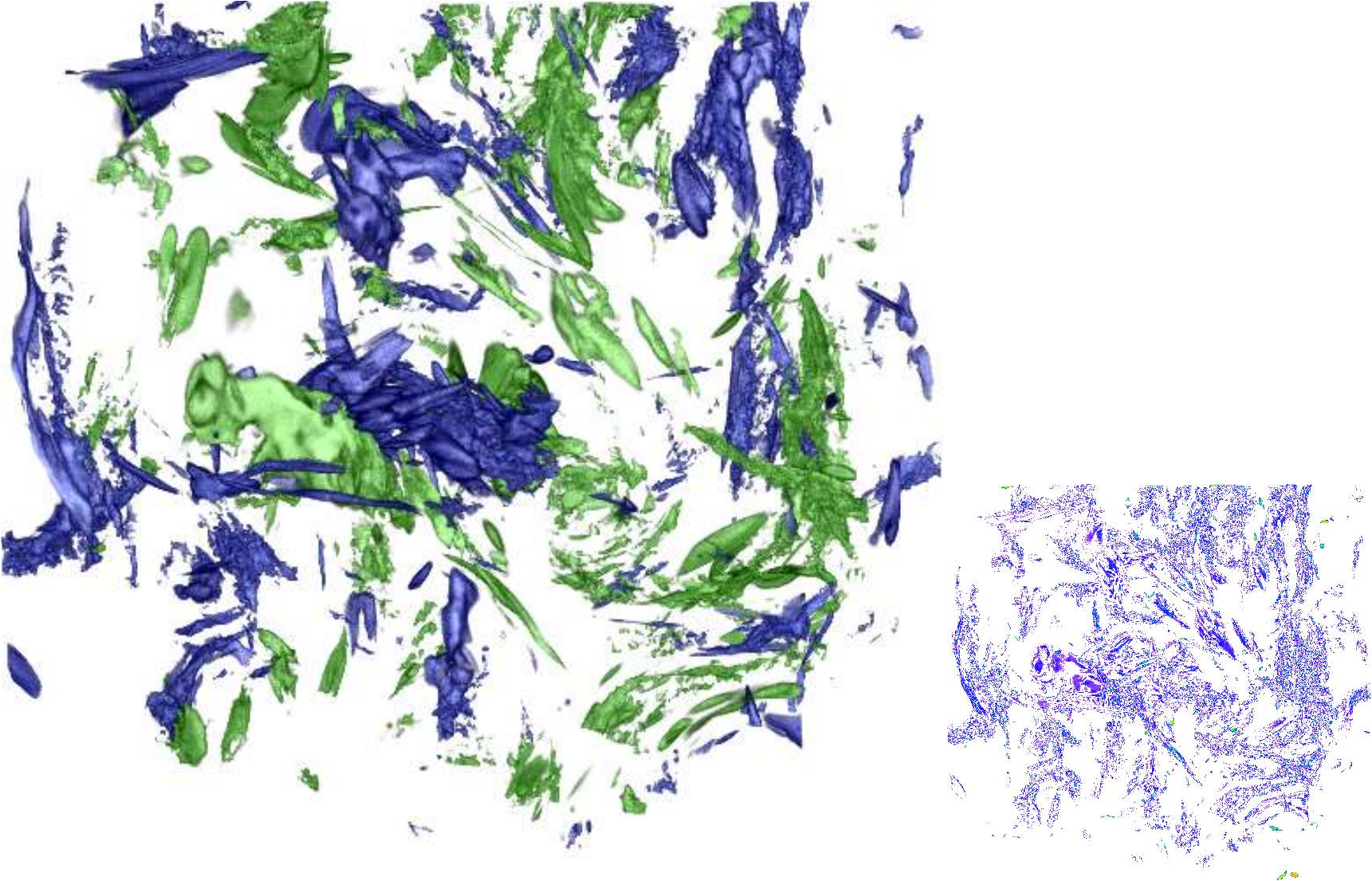}  \\
		\mbox{\footnotesize (c) E2}\\
\end{array}$
$\begin{array}{c@{\hspace{0.05in}}c}
		\includegraphics[width=0.45\linewidth]{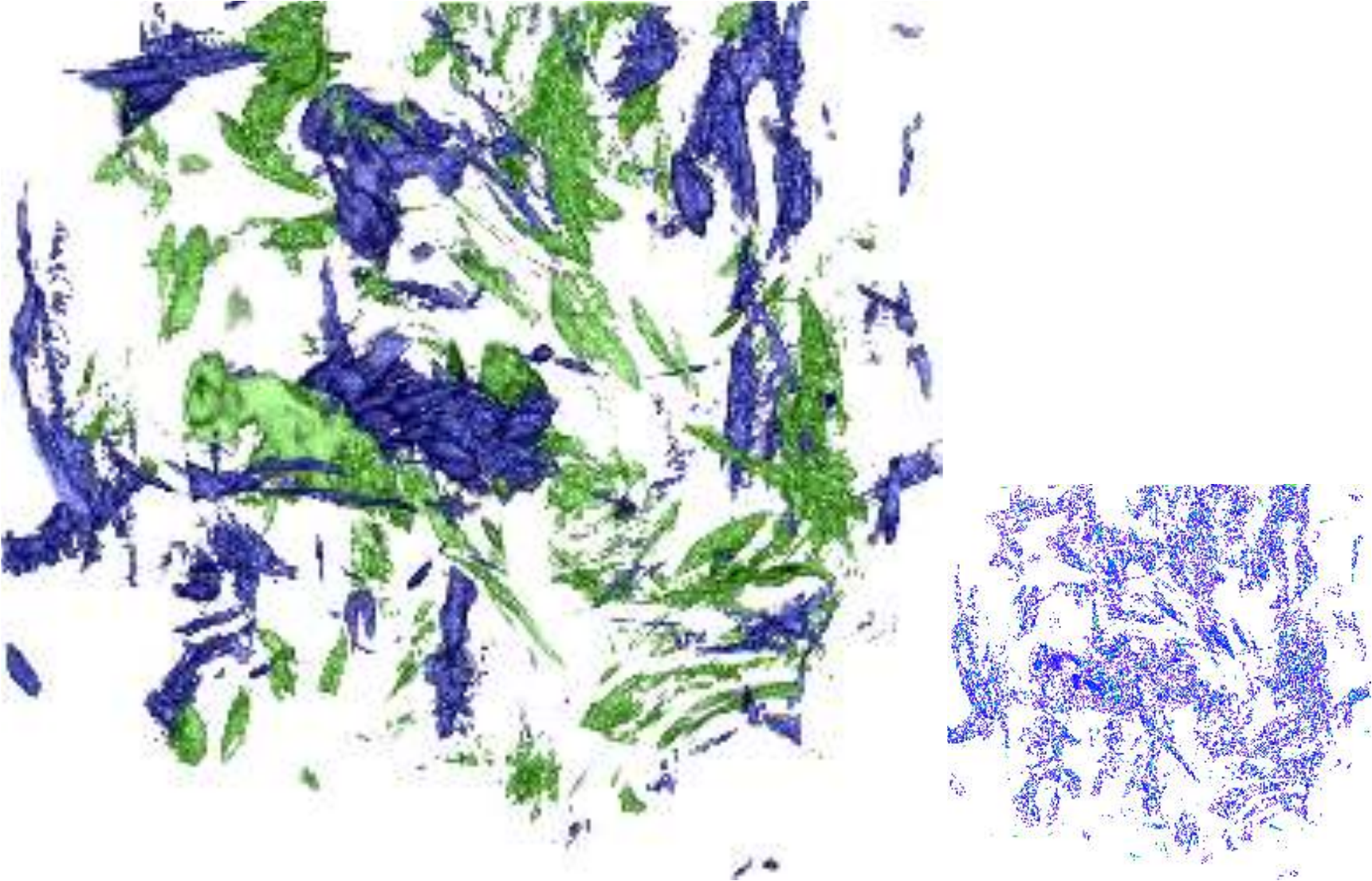}  &
		\includegraphics[width=0.45\linewidth]{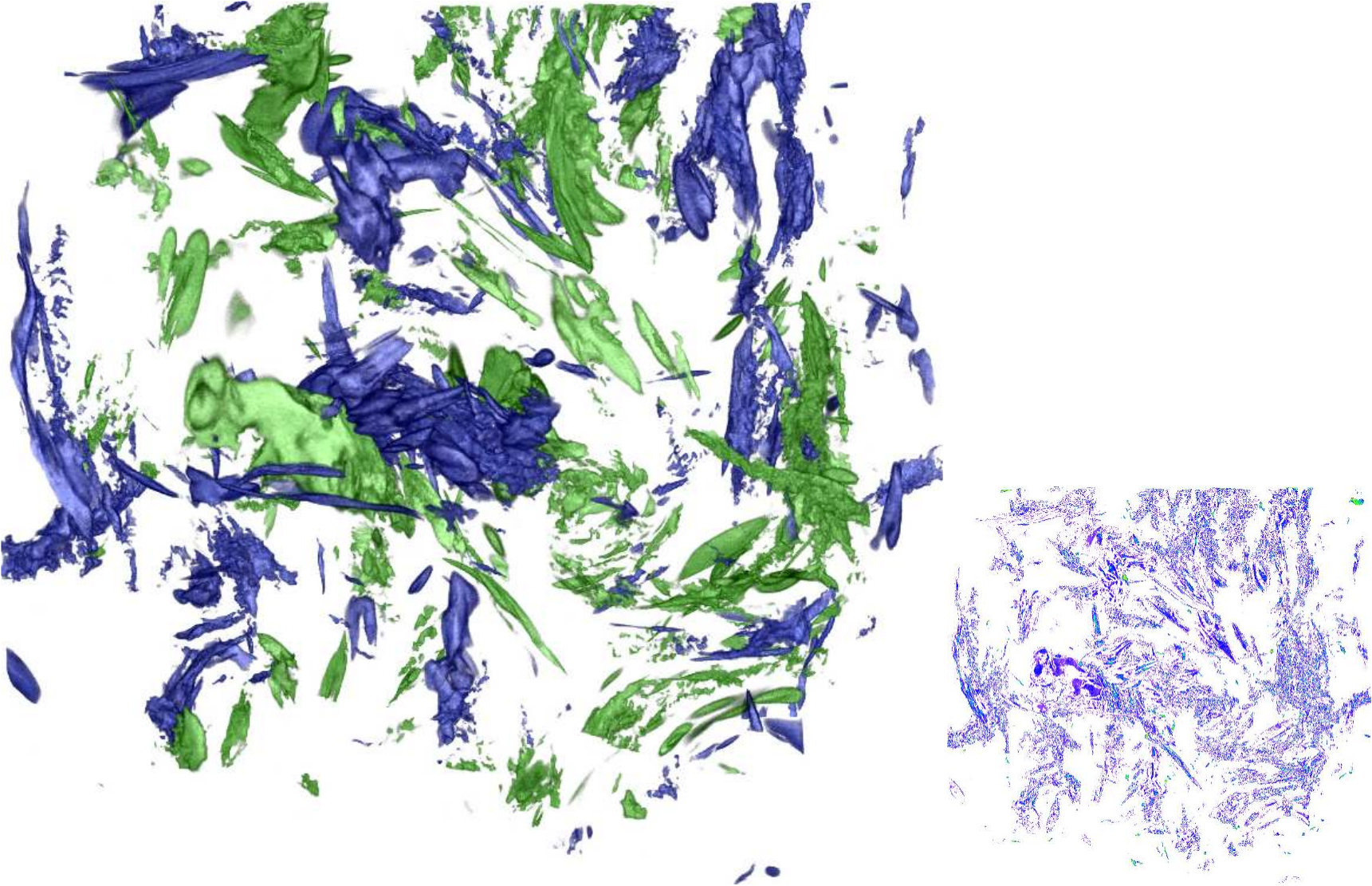}  \\
		\mbox{\footnotesize (d) E4} & \mbox{\footnotesize (e) E5}\\
\end{array}$
\end{center}
\vspace{-.25in}
\caption{\pin{PCN rendering results with different volume and image resolutions on the rotstrat dataset.}} 
\label{fig:PCN-vs-DVR}
\end{figure}

\pin{{\bf Comparison with NPR.}
Conventional {\em non-photorealistic rendering} (NPR) of a DVR scene relies on predefined generation rules to mimic a reference style. 
Due to the complexity of hyperparameter settings, the effectiveness of these methods depends on the choice of hyperparameters to some extent.
Moreover, such an approach limits the stylization results to a single style.
In contrast, StyleRF-VolVis can transfer arbitrary styles according to different reference images within a unified framework, offering higher flexibility and robustness.
In Figure~\ref{fig:Stipple}, we compare the stylization results of StyleRF-VolVis and a conventional NPR method for stipple drawing style~\cite{Lu-VIS02}.
We use the stipple drawing in~\cite{Lu-VIS02} as a reference image and adjust the TF and camera pose to align the viewpoint for comparison.
StyleRF-VolVis matches the overall texture of the reference style more closely.}

\begin{table*}[t]
\caption{\pin{Average rendering time, CPU memory, and GPU memory footprints for DVR and PCN as well as average PSNR (dB) and LPIPS values of PCN under different volume and image resolutions of the rotstrat dataset. Since E3 cannot be rendered in our local test machine using DVR due to out-of-memory (OOM), we record its performance on a high-performance cluster and highlight it in bold for reference.}}
\vspace{-0.1in}
\centering
{\scriptsize
\resizebox{\textwidth}{!}{
\begin{tabular}{c|cc|cccc|ccccc}
experiment&\multicolumn{2}{c|}{volume}  &\multicolumn{4}{c|}{DVR}		& \multicolumn{5}{c}{PCN} \\
ID&resolution	& size    		&resolution&time&CPU mem&GPU mem		&PSNR$\uparrow$ &LPIPS$\downarrow$ &time&CPU mem&GPU mem\\\hline
E1&$512^3$  	&0.5 GB    		&$800^2$&111 ms	&3.3 GB&1.4 GB	&28.71   &0.0326 &145 ms &4.9 GB&4.6 GB\\
E2&$1024^3$ 	&4.1 GB    		&$800^2$&388 ms	&11.3 GB&6.5 GB	&26.68   &0.0462 &153 ms&4.9 GB&4.6 GB\\
E3&$2048^3$ 	&32.8 GB   		&$800^2$&--/\textbf{3526 ms}	&OOM/\textbf{74.8 GB}&OOM/\textbf{45.5 GB}	&25.28   &0.0469 &166 ms&4.9 GB&4.6 GB\\ \hline
E4&$1024^3$  	&4.1 GB    		&$400^2$&323 ms	&11.2 GB&6.5 GB	&25.05   &0.0369 &76 ms&4.7 GB&3.1 GB\\
E2&$1024^3$ 	&4.1 GB    		&$800^2$&388 ms	&11.3 GB&6.5 GB	&26.68   &0.0462 &153 ms&4.9 GB&4.6 GB\\\
E5&$1024^3$  	&4.1 GB    		&$1200^2$&510 ms	&11.4 GB&	6.9 GB&26.89   &0.0503 &314 ms&5.2 GB&7.2 GB\\
\end{tabular}}
}
\label{tab:rotstrat-exp}
\end{table*}

\pin{{\bf Comparison with DVR.}
In DVR, users can adjust the TF to explore the scene. 
StyleRF-VolVis can also achieve similar objectives through the PSE of PCN. 
However, PSE cannot adjust invisible parts of the scene constructed via training the multiview images. 
Despite this limitation, PCN supports random access to any position within the scene during the rendering's sampling process. 
Compared to DVR, PCN may achieve faster render speed and require smaller memory footprints for large volumes under the same TF.
In Table~\ref{tab:rotstrat-exp}, we compare DVR and PCN regarding the average rendering time per image, CPU memory, and GPU memory requirement for the rotstrat dataset with different volume and image resolutions (denoted by experiment IDs). 
We run DVR using the open-source software ParaView with NVIDIA IndeX plugins.
All models converge around four minutes and occupy a storage of 168 MB.
PSNR and LPIPS values reported in Table~\ref{tab:rotstrat-exp} and the difference images (with respect to the GT rendering image) shown in Figure~\ref{fig:PCN-vs-DVR} suggest that PCN achieves acceptable accuracy but requires a smaller rendering memory footprint and faster rendering speed compared to DVR. 
Note that the NeRF-based representation is independent of the volume resolution.  
Storing the training images and the network model for large data is more space-efficient than the original volume. 
Therefore, StyleRF-VolVis provides an efficient means for altering renderings of large volumes.}

\end{document}